\documentclass[journal,draftcls,onecolumn,12pt,twoside]{IEEEtran}

\ifCLASSINFOpdf
   \usepackage[pdftex]{graphicx}
\else
\fi

\usepackage{hyperref}
\usepackage{bbm}
\usepackage{bm}
\usepackage{lipsum, color}
\usepackage{amsfonts}
\usepackage{amssymb}
 \usepackage{amsmath}
 \usepackage{cite}
\usepackage{pgfplots}
\usepackage[utf8]{inputenc}
\usepackage[english]{babel}

\usepackage{amsmath}  
\usepackage{amssymb}   
\usepackage{amsthm}  
\usepackage{amsmath}
\usepackage{stfloats}
\usepackage{blindtext}

\usepackage{cleveref}

\usepackage{lipsum}
\usepackage{mathtools}

\usepackage{cuted}
\usepackage{accents}

 \usepackage{booktabs}

\DeclareMathOperator*{\argmax}{arg\,max}  
  
  \usepackage{mathtools}
\usepackage{tikz}
\usetikzlibrary{shapes}

\newcommand\munderbar[1]{%
  \underaccent{\bar}{#1}}

 \newcommand{\Qu}[2][i]{\ensuremath{Q_{#1,#2}}} 
  \newcommand{\Es}[2][i]{\ensuremath{E_{s,#1#2}}} 
\newcommand{\Au}[2][i]{\ensuremath{\mathcal A_{#1}^{#2}}} 
 \newcommand{\Sf}[3][i]{\ensuremath{s_{\scalebox{1}{\tiny$#1#2#3$}}}} 
 \newcommand{\ro}[3][i]{\ensuremath{\rho_{\scalebox{1}{\tiny$#1#2#3$}}}}  
 \newcommand{\Pu}[3][i]{\ensuremath{p_{\scalebox{1}{\tiny$#1#2#3$}}}} 

\usepackage{colortbl}

\makeatletter
\newcommand{\vast}{\bBigg@{4}}
\newcommand{\Vast}{\bBigg@{5}}
\makeatother

\newtheorem{theorem}{Theorem}

\newtheorem{lemma}[theorem]{Lemma}
\newtheorem{proposition}{Proposition}

\pgfplotsset{width=5cm,compat=newest,every axis legend/.append style={at={(0.5,1.35)}, anchor=south}}

\begin{document}

\title{Multiple-Access Channel with Independent Sources: Error Exponent Analysis}




\author{Arezou Rezazadeh, Josep Font-Segura, Alfonso Martinez,\\ and Albert Guill{\'e}n i F{\`a}bregas
\thanks{A. Rezazadeh, J. Font-Segura and A. Martinez  are with the Department of Information and Communication Technologies, Universitat Pompeu Fabra, Barcelona 08018, Spain (e-mail: arezou.rezazadeh@upf.edu, \{josep.font-segura, alfonso.martinez\}@ieee.org).}
\thanks{A. Guill{\'e}n i F{\`a}bregas is with Instituci{\'o} Catalana de Recerca i Estudis Avan\c{c}ats, the Department of Information and Communication Technologies, Universitat Pompeu Fabra, Barcelona 08018, Spain, and also with the Department of Engineering, University of Cambridge, Cambridge CB2 1PZ, U.K. (e-mail: guillen@ieee.org).}
\thanks{This work has been funded in part by the European Research Council under grant 725411, and by the Spanish Ministry of Economy and Competitiveness under grant TEC2016-78434-C3-1-R.}
}

\maketitle

\begin{abstract}
In this paper, an achievable error exponent for the multiple-access channel with two independent sources is derived. For each user, the source messages are partitioned into two classes and codebooks are generated by drawing codewords from an input distribution depending on the class index of the source message. The partitioning thresholds that maximize the achievable exponent are given by the solution of a system of equations. We also derive both lower and upper bounds for the achievable exponent in terms of Gallager's source and channel functions. Finally, a numerical example shows that using the proposed ensemble gives a noticeable gain in terms of exponent with respect to independent identically distributed codebooks.  
\end{abstract} 
\IEEEpeerreviewmaketitle

\pagebreak 

\section{Introduction}
\label{part1}

For point-to-point communication, many studies show that 
 joint
source-channel coding might achieve a better error exponent than
 separate source-channel coding \cite{gala, Cs2, zhong, jscc}. One strategy for joint source-channel coding is to assign source messages to disjoint classes, and to use codewords generated according to a distribution that depends on the class index. This random-coding ensemble achieves the sphere-packing exponent in those cases where it is tight \cite{jscc}. 
 
 Recent studies \cite{farkas}, \cite{isit2017} extended the same idea to the  multiple-access channel (MAC) using a random-coding ensemble with independent message-dependent distributed codebooks. In \cite{isit2017}, the joint source-channel coding problem over a MAC with correlated sources was considered, 
where codewords are generated by a symbol-wise conditional probability
distribution  that depends both on the
instantaneous source symbol and on the empirical distribution of the source
sequence. The achievable exponent derived in \cite{isit2017} was presented in the primal domain, i.e., as a multi-dimensional optimization problem over distributions, which is generally difficult to analyze.

This work studies a simplified version of the problem posed in  \cite{isit2017} in the dual domain, i.e., as a lower dimensional problem over parameters in terms of Gallager functions. A two-user MAC with independent sources is considered.  For each user, source messages
are assigned to two classes, and
codewords are independently generated according to a distribution
that depends on the class index of the source message. For this random-coding ensemble, 
we derive an achievable exponent in the dual domain, and show that this exponent is greater than that achieved using only one input distribution for each user.

\section{System Model}

\subsection{Definitions and Notation}

We consider 
two independent sources characterized by probability distributions $P_{U_1}$, $P_{U_2}$ on alphabets $\mathcal U_1$ and $\mathcal U_2$, respectively. We use bold font to denote a sequence, such as the source sequences $\bm u_1\in{\cal U}_1^n$ and $\bm u_2\in{\cal U}_2^n$, and underlined font to represent a pair of quantities for users 1 and 2, such as $\munderbar {\gamma}= (\gamma_1,\gamma_2)$, $\munderbar {u}= (u_1,u_2)$, $\munderbar{\bm u} = (\bm u_1,\bm u_2)$ or $P_{\munderbar{ U}}(\munderbar u)=P_{U_1,U_2}(u_1,u_2)$.

For user $\nu=1,2$, the source message $\bm u_\nu$ is mapped onto codeword $\bm x_\nu(\bm u_\nu)$, which also has length $n$ and is drawn from the codebook $\mathcal{C}^\nu=\{\bm x_\nu(\bm u_\nu) \in {\cal X}_\nu^n \, : \, \bm u_\nu\in \mathcal{U}_\nu^n\}$.  
Both terminals send the codewords over a discrete memoryless multiple-access channel with transition probability $W(y|x_1,x_2)$, input alphabets $\mathcal{X}_1$ and $\mathcal{X}_2$, 
 and output alphabet $\mathcal{Y}$.

Given the received sequence $\bm y$, the decoder estimates the transmitted pair of messages $\munderbar{\bm u}$ based on the maximum a posteriori criterion, i.e.,
\begin{align}
\munderbar{\hat{\bm  u}}=\argmax_{\munderbar{\bm u} \, \in \, \mathcal{U}_1^n \times {\cal U}_2^n}P^n_{\munderbar{\bm U}}(\munderbar{\bm u})W^n\big(\bm y|\bm x_1(\bm u_1),\bm x_2(\bm u_2)\big).\label{criteria}
\end{align}
An error occurs if $\munderbar{\hat{\bm u}}\neq\munderbar{\bm u}$. Using the convention that scalar random variables are denoted by capital letters, e.g., $X$, and capital bold font letters denote random vectors, the error
probability for a given pair of codebooks $({\cal C}^1,{\cal C}^2)$ is given by
\begin{align}
\epsilon^n(\mathcal{C}^1,\mathcal{C}^2)\triangleq {\mathbb P}\left[(\hat{\bm{ U}}_1, \hat{\bm{U}}_2)\neq (\bm U_1, \bm U_2) \right].
\label{error2}
\end{align}

The pair of sources $(U_1,U_2)$ is transmissible over the channel if there exists a sequence of pairs of codebooks $(\mathcal{C}^1_n,\mathcal{C}^2_n)$ such that $\lim_{n\rightarrow\infty} \epsilon^n(\mathcal{C}^1_n,\mathcal{C}^2_n)=0$. An exponent $E(P_{\munderbar{U}},W)$ is achievable if there exists a sequence of codebooks such that
\begin{align}
 \liminf_{n\rightarrow \infty} -\dfrac{1}{n} \log \epsilon^n(\mathcal{C}^1_n,\mathcal{C}^2_n) \geq E(P_{\munderbar{U}},W).
\end{align}

In order to show the existence of such sequences of codebooks, we use random-coding arguments, i.e., 
we find a sequence of ensembles whose error probability averaged over the ensemble, denoted as $\bar{\epsilon}^n$, tends to zero.

\subsection{Message-Dependent Random Coding}

\label{sec:B}
For user $\nu=1,2$, we  fix a threshold $0\leq\gamma_\nu\leq 1$ to 
 partition the source-message set $\mathcal U_\nu^n$ into two classes $\Au[\nu]{1}$ and $\Au[\nu]{2}$ defined as
\begin{align}
\Au[\nu]{1}=\big\{\bm u_\nu\in \mathcal U^n_\nu \,:\, P_{\bm U_\nu}^n(\bm u_\nu)\geq \gamma_\nu^n\big\}\label{A1_itw2018},\\
\Au[\nu]{2}=\big\{\bm u_\nu\in \mathcal U^n_\nu \,:\, P_{\bm U_\nu}^n(\bm u_\nu)< \gamma_\nu^n\big\}.
\label{A2_itw2018}
\end{align}
For
every message
$\bm u_\nu \in \Au[\nu]{i}$, we randomly generate a codeword $\bm x_\nu(\bm u_\nu)$ according to the probability distribution $\Qu[\nu]{i}(\bm x_\nu)=\prod_{\ell=1}^n \Qu[\nu]{i}(x_{\nu,\ell})$, where $\Qu[\nu]{i}$, for $i=1,2$, is a probability distribution that depends on the class of $\bm u_\nu $.

We use the symbol $\tau\in\{\{1\},\{2\},\{1,2\}\}$ to denote the error event type of the error probability \eqref{error2}, i.e., respectively $(\hat{\bm u}_1,\bm u_2) \neq (\bm u_1, \bm u_2)$, $(\bm u_1,\hat{\bm u}_2) \neq (\bm u_1, \bm u_2)$ and $(\hat{\bm u}_1,\hat{\bm u}_2) \neq (\bm u_1, \bm u_2)$. We denote the complement of $\tau$ as $\tau^c$ among the subsets of $\{1,2\}$. For example, $\tau^c=\{2\}$ for $\tau=\{1\}$ and $\tau^c=\emptyset$ for $\tau=\{1,2\}$. In order to simplify some expressions, it will prove convenient to adopt the following notational convention for an arbitrary variable $u$
\begin{align}
u_{\tau}=\left\{
\begin{array}{ll}
\emptyset  & \tau=\emptyset\\
u_1  & \tau=\{1\}\\
u_2   & \tau=\{2\}\\
\munderbar u  & \tau=\{1,2\}.
\end{array} 
\right.
\end{align}

For types of error $\tau=\{1\}$ and $\tau=\{2\}$, we denote  $WQ_{\tau^c,i}$  as
 a point-to-point channel  
 with input and output alphabets given by $\mathcal X_\tau$ and  $\mathcal X_{\tau^c}\times \mathcal Y$, respectively, and transition probability $W(y|x_1,x_2)Q_{\tau^c,i}(x_{\tau^c})$. For $\tau=\{1,2\}$, the input distribution $Q_{\tau,i_\tau}=\Qu[1]{i_1}\Qu[2]{i_2}$ is the product distribution $\Qu[1]{i_1}(x_1)\Qu[2]{i_2}(x_2)$ over the alphabet $\mathcal X_1 \times \mathcal X_2$, and $WQ_{\tau^c,i}=W$,

\subsection{Single User Communication}
For point to point communication, using 
i.i.d random coding to
 transmit  a discrete memoryless source $P_U$, $u\in \mathcal U$ over the discrete memoryless channel $W$ with input and output  alphabets $\mathcal X$ and $\mathcal Y$,   leads to 
Gallager's  source and channel functions  \cite{gala}
\begin{align}
E_s(\rho,P_U)&=\log\Biggl(\sum_u  P_U(u)^{\frac{1}{1+\rho}}\Biggr)^{1+\rho}\label{Es_func},\\
E_0(\rho,Q,W)&=-\log\sum_y\Biggl( \sum_{x} Q(x)W(y|x)^{\frac{1}{1+\rho}}\Biggr)^{1+\rho}\label{E0_func},
\end{align}
where  $Q$ denotes the input distribution.

In \cite{jscc},  
message-dependent random coding was studied for single-user communication  using a threshold $\gamma\in[0,1]$ to partition the source messages into two classes. The derivation of the achievable exponent in \cite{jscc} involves the following  source exponent functions \cite[Lemma 1]{jscc}
\begin{align}
\Es[1]{}(\rho,P_{U},\gamma)=
\left\{
\begin{array}{r}
E_s(\rho,P_{U})\hspace{10.5em} \frac{1}{1+\rho} \geq\frac{1}{1+\rho_\gamma},
\\
E_s(\rho_\gamma,P_{U})+
E_s^\prime(\rho_\gamma)(\rho-\rho_\gamma)
\hspace{2.6em} 
\frac{1}{1+\rho} <\frac{1}{1+\rho_\gamma},
\end{array} \right.\label{Esi_itw_2018_1}
\end{align}
and
\begin{align}
\Es[2]{}(\rho,P_{U},\gamma)=
\left\{
\begin{array}{rl}
E_s(\rho,P_{U})\hspace{10.5em} \frac{1}{1+\rho} <\frac{1}{1+\rho_\gamma},\\
E_s(\rho_\gamma,P_{U})+E_s^\prime(\rho_\gamma)(\rho-\rho_\gamma)
\hspace{2.6em} \frac{1}{1+\rho} \geq\frac{1}{1+\rho_\gamma}.
\end{array} \right.\label{Esi_itw_2018_2}
\end{align}
In \eqref{Esi_itw_2018_1} and \eqref{Esi_itw_2018_2}, the parameter $\rho_\gamma$ is the solution of the implicit equation
\begin{align}
\frac{\sum_u P_U(u)^{\frac{1}{1+\rho}}\log P_U(u)}{\sum_u P_U(u)^{\frac{1}{1+\rho}}}=\log(\gamma),
\label{11}
\end{align}
when $\min_u P_U(u) \leq\gamma\leq \max_u P_U(u)$ is satisfied. We observe that $\Es[1]{}(\rho,\cdot)$ follows the Gallager $E_s(\rho,\cdot)$ function for an interval of $\rho$, while it is the straight line tangent to $E_s(\rho,\cdot)$ beyond that interval, and similarly for $\Es[2]{}(\rho,\cdot)$.

When $\gamma\in[0,\min_u P_U(u))$, we have that $\rho_\gamma=-1_-$ and hence $\Es[1]{}(\rho,\cdot)=E_s(\rho,\cdot)$ and $\Es[2]{}(\rho,\cdot)=-\infty$. Otherwise, when $\gamma\in(\max_u P_U(u),1]$, we have that $\rho_\gamma=-1_+$ and hence $\Es[1]{}(\rho,\cdot)=-\infty$ and $\Es[2]{}(\rho,\cdot)=E_s(\rho,\cdot)$. In our analysis, it suffices to consider $\gamma=0$ or $\gamma=1$ to represent the cases where $\Es[1]{}(\rho,\cdot)$ or $\Es[2]{}(\rho,\cdot)$ are infinity. For such cases, we have
\begin{align}
 \Es[1]{}(\rho,P_{U},0)&=E_s(\rho,P_{U}),\hspace{1em} \Es[2]{}(\rho,P_{U},0)=-\infty,\label{esi_for_gamma0}
\\
\Es[1]{}(\rho,P_{U},1)&=-\infty, \hspace{3.25em} \Es[2]{}(\rho,P_{U},1)=E_s(\rho,P_{U}).\label{esi_for_gamma1}
\end{align}

\section{Main Results}
\label{itw_result_sec}

We now derive an achievable exponent for the MAC with independent sources using the random-coding ensemble introduced in Sec. \ref{sec:B} in terms of the exponent functions defined in  \eqref{Es_func}--\eqref{Esi_itw_2018_2}. We also derive simpler lower and upper bounds to the achievable exponent in Sec. \ref{sec:lower} and \ref{sec:upper}, respectively.




\begin{proposition}
\label{pro1_itw_21.04}
For the two-user MAC with transition probability $W$,  source probability distributions $P_{\munderbar U}$ and class distributions $\{\Qu[\nu]{1}, \Qu[\nu]{2} \}$ with user index $\nu=1,2$, an achievable
exponent $E(P_{\munderbar U},W)$ is given by
\begin{align}
E(P_{\munderbar U},W)=\max_{\gamma_1,\gamma_2\in[0,1]}\min_{\tau\in\{\{1\},\{2\},\{1,2\} \}} \min_{i_{\scalebox{.8}{\tiny$\tau$}},i_{\scalebox{.8}{\tiny$\tau^c$}}=1,2} F_{\tau,i_\tau,i_{\tau^c}}(\gamma_1,\gamma_2),\label{our_exp_itw}
\end{align}
where
\begin{equation}
	F_{\tau,i_\tau,i_{\tau^c}}(\gamma_1,\gamma_2) = \max_{\rho\in[0,1]}
E_0(\rho,\Qu[\tau]{i_\tau},W\Qu[\tau^c]{i_{\tau^c}})
-\Es[i_\tau]{}(\rho,P_{ U_\tau},\gamma_\tau)-\Es[i_{\tau^c}]{}(0,P_{ U_{\tau^c}},\gamma_{\tau^c}).
\label{eq:F}
\end{equation}
In \eqref{eq:F}, the functions $E_0(\cdot)$, $\Es[1]{}(\cdot)$ and  $\Es[2]{}(\cdot)$ are respectively given by \eqref{E0_func}, \eqref{Esi_itw_2018_1} and \eqref{Esi_itw_2018_2}, and we define $\Es[i_{\{1,2\}}]{}(\rho,P_{\munderbar U},{\munderbar \gamma})=\Es[i_1]{}(\rho,P_{U_1},\gamma_1)+\Es[i_2]{}(\rho,P_{U_2},\gamma_2)$.
\end{proposition}
\begin{IEEEproof}
See Appendix \ref{itw-15:19-21.04.2018}.
\end{IEEEproof}

We remark that the optimal assignment of input distributions to source classes is considered in \eqref{our_exp_itw}. Since we considered two source-message classes $\Au[\nu]{1}$, $\Au[\nu]{2}$ and two input distributions $\Qu[\nu]{1}, \Qu[\nu]{2}$ for each user $\nu=1,2$, there are four possible assignments.

The derived achievable exponent \eqref{our_exp_itw} contains a maximization over $\gamma_1$ and $\gamma_2$, the thresholds that determine how source messages are partitioned into classes. Rearranging the minimizations over $\tau$, $i_\tau$ and $i_{\tau^c}$,  defining $f_{i_{1},i_{2}}( \gamma_1,\gamma_2)$ as
\begin{align}
f_{i_{1},i_{2}}( \gamma_1,\gamma_2)=\displaystyle\min_{\tau\in\{\{1\},\{2\},\{1,2\} \}} F_{\tau,i_\tau,i_{\tau^c}}(\gamma_1,\gamma_2),
\label{nef_al_itw_27_4}
\end{align}
where $F_{\tau,i_\tau,i_{\tau^c}}(\gamma_1,\gamma_2)$ is given in \eqref{eq:F}, the achievable exponent  \eqref{our_exp_itw}  can be written as
\begin{align}
E(P_{\munderbar U},W)=\max_{\gamma_1,\gamma_2\in[0,1]}\min_{i_1,i_2=1,2} f_{i_1,i_2}( \gamma_1,\gamma_2).\label{itw_13_32_4_19}
\end{align}

We note that regardless the values of $i_2$, $f_{1,i_{2}}(\munderbar \gamma)$ is non-decreasing with respect to $\gamma_1$ and
$f_{2,i_{2}}(\munderbar \gamma)$ is non-increasing with respect to $\gamma_1$. 
Similarly, regardless the values of $i_1$, $f_{i_{1},1}(\munderbar \gamma)$ is non-decreasing with respect to $\gamma_2$ and
$f_{i_{1},2}(\munderbar \gamma)$ is non-increasing with respect to $\gamma_2$. As a result, we derive a system of equations to compute the optimal thresholds $\gamma_1^\star$ and $\gamma_2^\star$.

 \begin{proposition}
  \label{itw_prop1}
 The optimal $\gamma_1^\star$ and $\gamma_2^\star$ maximizing  \eqref{our_exp_itw}  satisfy
 \begin{align}
 \left\{
 \begin{array}{rl}
\displaystyle\min_{i_{2}=1,2}
 f_{1,i_{2}}( \gamma_1^\star,\gamma_2^\star)=
\min_{i_{2}=1,2}
 f_{2,i_{2}}( \gamma_1^\star,\gamma_2^\star),
\\
\displaystyle \min_{i_{1}=1,2}
 f_{i_{1},1}( \gamma_1^\star,\gamma_2^\star)= \min_{i_{1}=1,2}
 f_{i_{1},2}( \gamma_1^\star,\gamma_2^\star) .\label{utw_optimal_gamma}
  \end{array}\right.
 \end{align}
When \eqref{utw_optimal_gamma} has no solutions, then  $\gamma_\nu^\star\in\{0,1\}$. In particular, if $ f_{1,i_{2}}(0, \gamma_2) > f_{2,i_{2}}(0, \gamma_2)$ then 
 $\gamma_1^\star=0$,  otherwise $\gamma_1^\star=1$; and 
 if $ f_{i_{1},1}(\gamma_1,0) > f_{i_{1},2}( \gamma_1,0)$, we have $\gamma_2^\star=0$, otherwise $\gamma_2^\star=1$.
\end{proposition}
\begin{IEEEproof}
See Appendix \ref{proof_prop2_itw}.
\end{IEEEproof}
 
We note that the optimal $\gamma_1^\star$ and $\gamma_2^\star$ are the points where the minimum of all non-decreasing functions with respect to $\gamma_\nu$ 
are equal with the minimum of all non-increasing functions with respect to $\gamma_\nu$, for both $\nu=1,2$. Even though $\gamma_1^\star$ and $\gamma_2^\star$ can be computed through equation \eqref{utw_optimal_gamma}, the final expression of the achievable exponent \eqref{our_exp_itw} is still coupled with $\gamma_1^\star$ and $\gamma_2^\star$. In the sequel, we alternatively study both lower and an upper bounds that do not depend on $\gamma_1$ and $\gamma_2$.

\subsection{A Lower Bound for the Achievable Exponent}

\label{sec:lower}

In order to find a lower  bound for the achievable exponent presented in \eqref{our_exp_itw}, we use properties \eqref{esi_for_gamma0} and \eqref{esi_for_gamma1}.
Firstly, we maximize over $\gamma_\nu\in\{0,1\}$ rather than $\gamma_\nu\in[0,1]$, for $\nu=1,2$, to lower bound  \eqref{our_exp_itw}. Let $d(\gamma_1,\gamma_2)$ be
\begin{equation}
	d(\gamma_1,\gamma_2)=\min_{i_1,i_2} f_{i_1,i_2}(\gamma_1,\gamma_2).\label{d_func_itw}
\end{equation}
Then,
\begin{equation}
	E(P_{\munderbar U},W) = \max_{\gamma_1,\gamma_2\in[0,1]}d(\gamma_1,\gamma_2) \geq \max_{\gamma_1,\gamma_2\in\{0,1\}} d(\gamma_1,\gamma_2).
\end{equation}
On the other hand,
\begin{align}
\max_{\gamma_1,\gamma_2\in\{0,1\}} d(\gamma_1,\gamma_2)
=\max\{d(0,0), d(0,1), d(1,0), d(1,1)\}.\label{sim_itw_24.4}
\end{align}
Taking into account properties \eqref{esi_for_gamma0} and \eqref{esi_for_gamma1},  we note that $f_{i_1,i_2}(\gamma_1,\gamma_2)$, for $\gamma_1,\gamma_2\in\{0,1\}$, is either infinity, or the Gallager's source-channel exponent, i.e., 
\begin{align}
\max_{\rho\in[0,1]}
E_0(\rho,\Qu[\tau]{i_\tau},W\Qu[\tau^c]{i_{\tau^c}})
-E_s(\rho,P_{ U_\tau}).
\label{name}
\end{align}
For example, $f_{i_1,i_2}(0,1)$ equals equation \eqref{name} for $i_1=1$ and $i_2=2$, and $f_{i_1,i_2}(0,1)=\infty$ for the rest of combinations of $i_1$ and $i_2$. Thus, $d(0,1)=\min_\tau\max_{\rho\in[0,1]}
E_0(\rho,\Qu[\tau]{i_\tau},W\Qu[\tau^c]{i_{\tau^c}})
-E_s(\rho,P_{ U_\tau})$ for $i_1=1$ and $i_2=2$. Similarly, $d(1,0)=\min_\tau\max_{\rho\in[0,1]}
E_0(\rho,\Qu[\tau]{i_\tau},W\Qu[\tau^c]{i_{\tau^c}})
-E_s(\rho,P_{ U_\tau})$ for $i_1=2$ and $i_2=1$, and so on. Hence, we obtain the following lower bound
\begin{align}
E(P_{\munderbar U},W)\geq E_{\rm L}(P_{\munderbar U},W),\label{pre_lb_itw}
\end{align}
where 
\begin{align}
E_{\rm L}(P_{\munderbar U},W)=\max_{i_1\in\{1,2\}} \max_{i_2\in\{1,2\}}
 \min_{\tau\in\{\{1\},\{2\},\{1,2\} \}} F^{\rm L}_{\tau,i_\tau,i_{\tau^c}},
\label{lb_itw}
\end{align}
with
\begin{equation}
	F^{\rm L}_{\tau,i_\tau,i_{\tau^c}} = \max_{\rho\in[0,1]}
E_0(\rho,\Qu[\tau]{i_\tau},W\Qu[\tau^c]{i_{\tau^c}})
-E_s(\rho,P_{ U_\tau}).
\label{eq:Fl}
\end{equation}

We note that for $\tau=\{1\}$ and $\tau=\{2\}$, $F^{\rm L}_{\tau,i_\tau,i_{\tau^c}}$ in \eqref{eq:Fl} is the error exponent of the point-to-point channel $W\Qu[\tau^c]{i_{\tau^c}}$ for an i.i.d. random-coding ensemble with distribution  $\Qu[\tau]{i}$. For $\tau=\{1,2\}$, we have $W\Qu[\tau^c]{i_{\tau^c}}=W$ and $E_s(\rho,P_{ U_\tau})=E_s(\rho,P_{ U_1})+E_s(\rho,P_{ U_2})$, so that \eqref{eq:Fl} is the error exponent of the point-to-point channel $W$ for an i.i.d. random-coding ensemble with distribution $\Qu[1]{i_1}\Qu[2]{i_2}$. Hence, the lower bound \eqref{lb_itw} selects the best assignment of input distributions over all four combinations through $i_1$ and $i_2$.

\subsection{An Upper Bound for the Achievable Exponent}

\label{sec:upper}

Now, we derive an upper bound for \eqref{our_exp_itw}
inspired by the tools used in \cite{jscc} for single user communication. Let $E_0(\rho,\mathcal Q,W)=\max_{Q\in\mathcal Q} E_0(\rho,Q,W)$, where ${\cal Q}$ is a set of distributions. We
 denote $\bar E_0(\rho,\mathcal Q,W)$ as the concave hull of $E_0(\rho,\mathcal Q,W)$, defined as the point-wise supremum over all 
convex combinations of any two values of the function $E_0(\rho,\mathcal Q,W)$, i.e.,
\begin{align}
\bar E_0(\rho,\mathcal Q,W)
\triangleq \sup_{\substack{\rho_1,\rho_2,\theta\in[0,1] \,:
\\
\theta\rho_1+(1-\theta)\rho_2=\rho} }\Bigl\{
\theta E_0(\rho_1,\mathcal Q,W)
+
(1-\theta) E_0(\rho_2,\mathcal Q,W)
\Bigr\}.\label{def_concave_huul_itw}
\end{align}
In \cite{jscc}, it is proved that joint source-channel random coding where source messages are assigned to different classes and codewords are generated according to a distribution that depends on the  class index of source message, achieves the following exponent
\begin{align}
\max_{\rho\in[0,1]}\bar E_0(\rho,\mathcal Q,W)
-E_s(\rho,P_U),\label{needed_ttw_p2p}
\end{align}
which coincides with the sphere-packing exponent \cite[Lemma 2]{Cs2} whenever it is tight.

For the MAC with independent sources, we use the max-min inequality \cite{fans_minimax} to upper-bound \eqref{our_exp_itw} by
 swapping the maximization over $\gamma_1$,$\gamma_2$ 
 with the minimization over $\tau$. Then, for a given $\tau$, we use Lemma \ref{itw_lema3} in Appendix \ref{proof_pre_up_bound_itw} to obtain the following result.
 \begin{proposition}
 The achievable exponent \eqref{our_exp_itw} is upper bounded as
\begin{align}
E(P_{\munderbar U},W)\leq E_{\rm U}(P_{\munderbar U},W),\label{pre_up_bound_itw}
\end{align}
where
\begin{align}
E_{\rm U}(P_{\munderbar U},W)=\min_{\tau\in\{\{1\},\{2\},\{1,2\} \}} F^{\rm U}_{\tau},
\label{upb_itw}
\end{align}
where
\begin{equation}
	F^{\rm U}_{\tau} = \max_{i_{\tau^c}=1,2}
\max_{\rho\in[0,1]} \bar E_0(\rho, \{\Qu[\tau]{1},\Qu[\tau]{2}\},W\Qu[\tau^c]{i_{\tau^c}})-E_s(\rho,P_{ U_\tau}).
	\label{eq:Fu}
\end{equation}
We recall that for $\tau=\{1,2\}$, we have $\{\Qu[\tau]{1},\Qu[\tau]{2}\}=\{\Qu[1]{1},\Qu[2]{1},\Qu[1]{2},\Qu[2]{2}\}$ and $E_s(\rho,P_{ U_\tau})=E_s(\rho,P_{ U_1})+E_s(\rho,P_{ U_2})$.
 \end{proposition}
\begin{IEEEproof}
See Appendix \ref{proof_pre_up_bound_itw}.
\end{IEEEproof}

From equation \eqref{upb_itw}, we observe that the upper bound is the minimum of three terms depending on $\tau\in\{\{1\},\{2\},\{1,2\} \}$. For $\tau\in\{\{1\},\{2\}\}$, we know that the message of user $\tau^c$ is decoded correctly so that user $\tau$ is virtually sent either over channel $W\Qu[\tau^c]{1}$ or $W\Qu[\tau^c]{2}$. Hence, the objective function of \eqref{upb_itw} is the single-user exponent for source $P_{ U_\tau}$
and point-to-point channel $W\Qu[\tau^c]{i_{\tau^c}}$ where codewords are generated according to two assigned input distributions $\{\Qu[\tau]{1},\Qu[\tau]{2}\}$ depending on class index of source messages. As a result, we note that the maximization over $i_{\tau^c}=1,2$ is equivalent to choose the best channel (either $W\Qu[\tau^c]{1}$ or $W\Qu[\tau^c]{2}$) in terms of error exponent.

\subsection{Numerical Example}
 \label{num_example_itw}
 Here we provide a numerical example comparing the achievable exponent, the lower bound and the upper bound  given in \eqref{our_exp_itw},  \eqref{lb_itw} and \eqref{upb_itw}, respectively.
 We consider
 two independent discrete memoryless sources with alphabet $\mathcal U_\nu=\{1,2\}$ for $\nu=1,2$ where $P_{U_1}(1)=0.028$ and $P_{U_2}(1)=0.01155$. We also consider
 a discrete memoryless  multiple-access channel with $\mathcal{X}_1=\mathcal{X}_2=\{1,2,\ldots,6\}$ and $|\mathcal{Y}|=4$.
 The transition probability of this channel, denoted as $W$, 
 is given by
  \begin{align}
    W=\left (
     \begin{array}{cl}
     W_{1}\\
     W_2\\
     W_3\\
     W_4\\
     W_5\\
   W_6
     \end{array}\right), \label{W_itw_final}
 \end{align}
 where 
 \begin{align}
W_{1}=\left( \begin{matrix} 
1-3k_1& k_1&k_1&k_1\cr
k_1&1-3k_1& k_1&k_1\cr
k_1&k_1&1-3k_1& k_1\cr
k_1&k_1& k_1&1-3k_1\cr
0.5-k_2& 0.5-k_2& k_2& k_2\cr
 k_2& k_2&0.5-k_2& 0.5-k_2
\end{matrix} \right),
 \end{align}
 for $k_1=0.056$ and $k_2=0.01$. $W_2$ and $W_3$ are $6\times4$ matrices whose rows are all the copy of $5^{\text{th}}$ and $6^{\text{th}}$ row of matrix $W_1$, respectively. Let the 
$m$-th row of matrix $W_1$ is denoted by $W_1(m)$. $W_4$, $W_5$ and $W_6$ are respectively given by
\begin{align}
    W_4=\left (
     \begin{array}{cl}
     W_{1}(2)\\
     W_1(3)\\
     W_1(4)\\
     W_1(1)\\
     W_1(6)\\
   W_1(5)
     \end{array}\right) 
     \hspace{1em}W_5=\left (
     \begin{array}{cl}
     W_{1}(3)\\
     W_1(4)\\
     W_1(1)\\
     W_1(2)\\
     W_1(5)\\
   W_1(6)
     \end{array}\right)\hspace{1em} W_6=\left (
     \begin{array}{cl}
     W_{1}(4)\\
     W_1(1)\\
     W_1(2)\\
     W_1(3)\\
     W_1(6)\\
   W_1(5)
     \end{array}\right).
\end{align}

We observe that $W$ is a $36\times 4$ matrix where the transition probability  $W(y|x_1,x_2)$ is placed at the row  $x_1+6(x_2-1)$ of matrix $W$, for
 $(x_1,x_2)\in\{1,2,...,6\}\times\{1,2,...,6\}$.
Recalling that each source has two classes and that four input distributions generate codewords, there are four possible assignments of input distributions to classes. 
 Among all possible permutations, we select the one that gives the highest exponent. Here, 
for user $\nu=1,2$, we consider the set of input distributions  $\big\{[0\hspace{.4em}0 \hspace{.4em}0\hspace{.4em}0\hspace{.4em}0.5
 \hspace{.4em}0.5],$ $[0.25\hspace{.4em}0.25 \hspace{.4em}0.25\hspace{.4em} 0.25 \hspace{.4em}0\hspace{.4em}0]\big\}$. 
For the channel given in \eqref{W_itw_final}, 
 the optimal assignment is
 \begin{align}
 \Qu[\nu]{1}=[0\hspace{.4em}0 \hspace{.4em}0\hspace{.4em}0\hspace{.4em}0.5
 \hspace{.4em}0.5],\hspace{3.5em} \label{eq:q1}\\
 \Qu[\nu]{2}=[0.25\hspace{.4em}0.25 \hspace{.4em}0.25\hspace{.4em} 0.25 \hspace{.4em}0\hspace{.4em}0],\label{eq:q2}
 \end{align}
for both $\nu=1,2$. Since we consider two input distributions for each user, the function $\max_{\rho\in[0,1]}$ 
$E_0(\rho,\Qu[\tau]{i_\tau},W\Qu[\tau^c]{i_{\tau^c}})$ is not concave in $\rho$ \cite{jscc}.
 For this example, from \eqref{utw_optimal_gamma},
we numerically  compute the optimal $\gamma_1^\star$ and $\gamma_2^\star$  maximizing \eqref{our_exp_itw} leading to $\gamma_1^\star=0.8159$ and  $\gamma_2^\star=0.7057$.

\begin {table}[!t]
\begin{center}
\caption{ Values of $F_{\tau,i_\tau,i_{\tau^c}}(\gamma_1^\star,\gamma_2^\star)$ in \eqref{eq:F} with optimal thresholds $\gamma_1^\star=0.8159$  $\gamma_2^\star=0.7057$, for types of error $\tau$, and user classes $i_\tau$ and $i_{\tau^c}$.}
\label{tab1_itw}            
\begin{tabular}{*{5}{c}}
\toprule
\multicolumn{5}{c}{$\hspace{6em}(i_1,i_2)$}
\\
\cline{2-5}
&(1,1)&(2,1)&(1,2)&(2,2)
\\
\midrule
 $\tau=\{1\}$&0.2566 &0.1721 &\cellcolor[gray]{.9}
{0.1057} & 0.1103
\\
 $\tau=\{2\}$&0.2597&\cellcolor[gray]{0.9}{0.1057}&0.2526& 0.2087
\\
 $\tau=\{1,2\}$& \cellcolor[gray]{0.9}{0.1057}&0.1073& 0.1127&0.1180
\\
\bottomrule
\end{tabular}
\end{center}
\end {table}
\begin {table}[!t]
\begin{center}  
\caption{Values of $F^{\rm L}_{\tau,i_\tau,i_{\tau^c}}$ in \eqref{eq:Fl} for types of error $\tau$, and input distribution $\Qu[1]{i_1},\Qu[2]{i_2}$.}
\label{tab2_itw}    
\begin{tabular}{*{5}{c}} 
\toprule
&\Qu[1]{1},\Qu[2]{1}&\Qu[1]{2},\Qu[2]{1}&\Qu[1]{1},\Qu[2]{2}&\Qu[1]{2},\Qu[2]{2}
\\
\midrule
 $\tau=\{1\}$&0.1723&0.1721&\cellcolor[gray]{0.9}0.0251&\cellcolor[gray]{0.9}0.0342
\\
 $\tau=\{2\}$&0.2526&\cellcolor[gray]{0.9}0.0989&0.2526& 0.2019
\\
 $\tau=\{1,2\}$&\cellcolor[gray]{0.9}0.0900&0.1073&0.0900&  0.0984
 \\
\bottomrule
\end{tabular}
\end{center}
\end {table}
\begin {table}[!t]
\begin{center}
\caption{Values of $F^{\rm U}_{\tau}$ in \eqref{eq:Fu} for types of error $\tau$.}
\label{tab3_itw}
\begin{tabular}{*{3}{c}} 
\toprule
$\tau=\{1\}$&$\tau=\{2\}$&$\tau=\{1,2\}$
\\
\midrule
0.1734 &0.2526&\cellcolor[gray]{0.9}0.1073
 \\
\bottomrule
\end{tabular}
\end{center}
\end {table}

Tables \ref{tab1_itw}, \ref{tab2_itw} and \ref{tab3_itw} respectively show  the objective functions $F_{\tau,i_\tau,i_{\tau^c}}(\gamma_1,\gamma_2)$, $F^{\rm L}_{\tau,i_\tau,i_{\tau^c}}$, and $F^{\rm U}_{\tau}$ given in \eqref{eq:F}, \eqref{eq:Fl} and \eqref{eq:Fu}, involved in the derivation of the achievable exponent  \eqref{our_exp_itw}, lower bound \eqref{lb_itw} and  upper bound \eqref{upb_itw}. The shaded elements in Tables  \ref{tab1_itw} and \ref{tab3_itw} respectively are the exponent and the upper bound. Additionally, the shaded  elements in Table \ref{tab2_itw} are the i.i.d.  exponent for different input distributions  assignments. Solving equations \eqref{our_exp_itw},  \eqref{lb_itw},  \eqref{upb_itw} using the partial optimizations in Tables  \ref{tab1_itw}, \ref{tab2_itw} and \ref{tab3_itw}, we respectively obtain 
 \begin{align}
 E(P_{\munderbar U},W)=0.1057,\\
 E_{\rm L}(P_{\munderbar U},W)=0.0989,\\
 E_{\rm U}(P_{\munderbar U},W)=0.1073.
 \end{align}

We observe that the percentage difference between the achievable exponent $E(P_{\munderbar U},W)$ and the lower bound $E_{\rm L}(P_{\munderbar U},W)$ is $6.875\% $. For a given set of two distributions for each user, the lower bound $E_{\rm L}(P_{\munderbar U},W)$ corresponds to the i.i.d. random-coding error exponent when each user uses only one input distribution. In \cite{jscc}, a similar comparison is made for point-to-point communication where the exponent achieved by an ensemble with two distributions is $0.75\%$ higher than the one achieved by the i.i.d. ensemble. Hence, our example illustrates that using message-dependent random coding with two class distributions may lead to higher error exponent gain in the MAC than in point-to-point communication, compared to i.i.d. random coding.




\appendices
\section{Proof of Proposition
\ref{pro1_itw_21.04}}
\label{itw-15:19-21.04.2018}
In order to prove Proposition \ref{pro1_itw_21.04}, we follow similar steps than in \cite{jscc}. Firstly, we start
 by bounding  $\bar \epsilon^n$, the  average error probability over the ensemble, for a given block length $n$. 
Applying the random-coding union bound \cite{polya} for joint source-channel coding, we have
\begin{align}
\bar \epsilon ^n\leq  \sum_{\munderbar {\bm u}, \munderbar {\bm x}, \bm y}P^n_{\munderbar{\bm U} \munderbar{\bm X} \bm Y}(\munderbar{\bm u}, \munderbar{\bm x},\bm y)
\min\Bigg\{1, \sum_{\substack{\munderbar {\hat{\bm u}} \neq \munderbar{\bm u} }} {\mathbb P}\bigg[\dfrac{P^n_{\munderbar{ \bm U}}(\munderbar{ \hat{\bm u}})W^n(\bm {y}|\munderbar{\hat{ \bm X}})}{P^n_{\munderbar{ \bm U}}(\munderbar{ \bm u})W^n(\bm {y}|\munderbar{ \bm x}) }\geq 1 \bigg]\Bigg\},
\end{align}
where $\munderbar{ \hat{\bm x}}$ has the same distribution as $\munderbar{ \bm x}$ but is independent of $\bm y$. 
The summation over $\munderbar{\hat{\bm u}} \neq \munderbar{\bm u}$ can be grouped into 
three types of error events, specifically $(\hat{\bm u}_1,\bm u_2) \neq (\bm u_1, \bm u_2)$, $(\bm u_1,\hat{\bm u}_2) \neq (\bm u_1, \bm u_2)$ and $(\hat{\bm u}_1,\hat{\bm u}_2) \neq (\bm u_1, \bm u_2)$. These three types of error events are denoted by $\tau\in\{ \{1\},\{2\},\{1,2\}\}$,  respectively. 
Using the fact that $\min\{1,a+b\}\leq\min\{1,a\}+\min\{1,b\}$, we further bound $\bar \epsilon^n$ as
\begin{align}
\bar \epsilon^n\leq \sum_\tau\bar \epsilon^n_\tau,\label{xi_first}
\end{align}
where
 \begin{align}
 \bar{\epsilon}_\tau^n\leq 
 \sum_{\munderbar{\bm u}}P^n_{\munderbar{\bm U} }(\munderbar{\bm u})
 \sum_{\munderbar{\bm x}, \bm y}P^n_{\munderbar{\bm X} \bm Y}( \munderbar{\bm x},\bm y)
\min\vast\{1, 
 \sum_{\hat{\bm u}_{\scalebox{.8}{\tiny$\tau$}}\neq \bm u_{\scalebox{.8}{\tiny$\tau$}}}\sum_{
 \hat{\bm  x }_\tau:\frac{P^n_{\scalebox{.8}{\tiny$\bm U$}_{\scalebox{.8}{\tiny$\tau$}}}(\hat{\bm{u}}_{\scalebox{0.8}{\tiny$\tau$}})W^n(\bm {y}|\hat{\bm x}_{\scalebox{0.8}{\tiny$\tau$}},\bm x_{\scalebox{0.8}{\tiny$\tau^c$}})}{P^n_{\scalebox{.8}{\tiny$\bm U$}_{\scalebox{.8}{\tiny$\tau$}}}(\bm{u}_{\scalebox{0.8}{\tiny$\tau$}})W^n(\bm {y}|\bm x_{\scalebox{0.8}{\tiny$1$}},\bm x_{\scalebox{0.8}{\tiny$2$}}) }
\geq 1}
\Qu[\tau]{\hat{\bm u}_\tau}^n(\hat{\bm{x}}_\tau)\vast\},\label{itw1}
 \end{align}
and $\Qu[\tau]{\hat{\bm u}_\tau}^n$
denotes the channel-input distribution corresponding
to the source message $\hat{\bm u}_{\scalebox{1}{\tiny$\tau$}}$.

Next, we break the summation over $\munderbar{\bm u}$  in \eqref{itw1} into the summations over the messages belonging to the classes $\Au[\nu]{1}$, $\Au[\nu]{2}$ and then summed over all classes. Moreover, by considering the case where codewords are generated according to distributions
that depend on the class index of the sources, the outer summation  of \eqref{itw1},  can be written as
 \begin{align}
\sum_{\munderbar{\bm u}}P^n_{\munderbar{\bm U} }(\munderbar{\bm u})
 \sum_{\munderbar{\bm x}, \bm y}P^n_{\munderbar{\bm X} \bm Y}( \munderbar{\bm x},\bm y)= 
\sum_{i_{\scalebox{.8}{\tiny$1$}},i_{\scalebox{.8}{\tiny$2$}}=1,2}
\sum_{\bm u_{\scalebox{.8}{\tiny$1$}}\in \Au[1]{i_1}}
P^n_{\bm U_1}(\bm u_1)\sum_{\bm u_{\scalebox{.8}{\tiny$2$}}\in \Au[2]{i_2}}
P^n_{\bm U_2}(\bm u_2)
\hspace{8em}\nonumber\\
\times\sum_{\munderbar{\bm x},\bm y}\Qu[1]{i_{\scalebox{.8}{\tiny$1$}}}^n(\bm x_1)
 \Qu[2]{i_{\scalebox{.8}{\tiny$2$}}}^n(\bm x_2)
 W^n(\bm y|\bm x_1,\bm x_2). \label{itw2}
\end{align}

Similarly, the inner summation of \eqref{itw1} 
can be grouped based on the classes of $\hat{\bm u}_{\scalebox{1}{\tiny$\tau$}}$ and then sum over all classes. Applying this fact and 
in view of
Markov's inequality for $\Sf[i_\tau]{}{j_\tau}\geq 0$, the inner summation of \eqref{itw1} is bounded as
 \begin{align}
\sum_{\hat{\bm u}_{\scalebox{.8}{\tiny$\tau$}}\neq \bm u_{\scalebox{.8}{\tiny$\tau$}}}\sum_{
\hat{\bm  x }_\tau:\frac{P^n_{\scalebox{1}{\tiny$\bm U$}_{\scalebox{.8}{\tiny$\tau$}}}(\hat{\bm{u}}_{\scalebox{0.8}{\tiny$\tau$}})W^n(\bm {y}|\hat{\bm x}_{\scalebox{0.8}{\tiny$\tau$}},\bm x_{\scalebox{0.8}{\tiny$\tau^c$}})}{P^n_{\scalebox{1}{\tiny$\bm U$}_{\scalebox{.8}{\tiny$\tau$}}}(\bm{u}_{\scalebox{0.8}{\tiny$\tau$}})W^n(\bm {y}|\bm x_{\scalebox{0.8}{\tiny$1$}},\bm x_{\scalebox{0.8}{\tiny$2$}}) }
\geq 1 } \Qu[\tau]{j_\tau }^n(\hat{\bm{x}}_\tau)\leq
\sum_{j_{\scalebox{.8}{\tiny$\tau$}}=1,2}
 \sum_{\hat{\bm u}_{\scalebox{.8}{\tiny$\tau$}}
 \in \Au[\tau]{j_\tau}}\sum_{\hat{\bm  x}_\tau}
\Qu[\tau]{j_\tau }^n(\hat{\bm{x}}_\tau)
\left(\frac{P^n_{\scalebox{1}{\tiny$\bm U$}_{\scalebox{.8}{\tiny$\tau$}}}(\hat{\bm{u}}_{\scalebox{0.8}{\tiny$\tau$}})W^n(\bm {y}|\hat{\bm x}_{\scalebox{0.8}{\tiny$\tau$}},\bm x_{\scalebox{0.8}{\tiny$\tau^c$}})}{P^n_{\scalebox{1}{\tiny$\bm U$}_{\scalebox{.8}{\tiny$\tau$}}}(\bm{u}_{\scalebox{0.8}{\tiny$\tau$}})W^n(\bm {y}|\bm x_{\scalebox{0.8}{\tiny$1$}},\bm x_{\scalebox{0.8}{\tiny$2$}}) } \right)^{\Sf[i_\tau]{}{j_\tau}}.
 \label{itw3}
 \end{align}
Inserting \eqref{itw3} into the inner minimization of \eqref{itw1} and using
 the inequality $\min\{1,A+B\}\leq \min_{\rho,\rho^\prime\in[0,1]}A^\rho+B^{\rho^\prime}$ for $A,B\geq 0$, $\rho,\rho^\prime \in [0,1]$, the inner term of  \eqref{itw1} is derived as
\begin{align}
\min\left\{1,\sum_{\hat{\bm u}_{\scalebox{.8}{\tiny$\tau$}}\neq \bm u_{\scalebox{.8}{\tiny$\tau$}}}\sum_{
\hat{\bm  x}_\tau:\frac{P^n_{\scalebox{.8}{\tiny$\bm U$}_{\scalebox{.8}{\tiny$\tau$}}}(\hat{\bm{u}}_{\scalebox{0.8}{\tiny$\tau$}})W^n(\bm {y}|\hat{\bm x}_{\scalebox{0.8}{\tiny$\tau$}},\bm x_{\scalebox{0.8}{\tiny$\tau^c$}})}{P^n_{\scalebox{.8}{\tiny$\bm U$}_{\scalebox{.8}{\tiny$\tau$}}}(\bm{u}_{\scalebox{0.8}{\tiny$\tau$}})W^n(\bm {y}|\bm x_{\scalebox{0.8}{\tiny$1$}},\bm x_{\scalebox{0.8}{\tiny$2$}}) }
\geq 1 } \Qu[\tau]{j_\tau }^n(\hat{\bm{x}}_\tau)\right\}
\hspace{16em}\nonumber
\\
\leq\sum_{j_{\scalebox{.8}{\tiny$\tau$}}=1,2}
\min_{\ro[i_\tau]{}{j_\tau}\in [0,1]} \frac{G_{j_\tau}(\Sf[i_\tau]{}{j_\tau},\bm x_{\tau^c},\bm y)^{\ro[i_\tau]{}{j_\tau}}}{\left(P^n_{\scalebox{1}{\tiny$\bm U$}_{\scalebox{.8}{\tiny$\tau$}}}(\bm{u}_{\scalebox{0.8}{\tiny$\tau$}})W^n(\bm {y}|\bm x_{\scalebox{0.8}{\tiny$1$}},\bm x_{\scalebox{0.8}{\tiny$2$}})\right)
 ^{\Sf[i_\tau]{}{j_\tau}\ro[i_\tau]{}{j_\tau}}
 }
,\label{itw4}
 \end{align}
 where
 \begin{align}
G_{i_\tau}(s,\bm x_{\tau^c},\bm y)=
\sum_{\substack{\bm u_{\scalebox{.8}{\tiny$\tau$}}
 \in \Au[\tau]{i_\tau}\\
 \bm  x_\tau
}}
P^n_{\scalebox{1}{\tiny$\bm U$}_{\scalebox{.8}{\tiny$\tau$}}}(\bm{u}_{\scalebox{0.8}{\tiny$\tau$}})^s\Qu[\tau]{i_\tau }^n(\bm{x}_\tau)
W^n(\bm {y}|\bm x_{\scalebox{0.8}{\tiny$\tau$}},\bm x_{\scalebox{0.8}{\tiny$\tau^c$}})^s,\label{G_def}
\end{align}
and $\ro[i_\tau]{}{j_\tau}\in [0,1]$ and 
$\Sf[i_\tau]{}{j_\tau}\geq 0$.
By putting back \eqref{itw2} and \eqref{itw4} into the respective outer and inner terms of \eqref{itw1}, the average error probability is bounded as
 \begin{align}
 \bar{\epsilon}_\tau^n\leq 
\sum_{j_{\scalebox{.8}{\tiny$\tau$}}=1,2}\sum_{i_{\scalebox{.8}{\tiny$1$}},i_{\scalebox{.8}{\tiny$2$}}=1,2}
\min_{\ro[i_{\scalebox{.8}{\tiny$\tau$}}]{}{j_{\scalebox{.8}{\tiny$\tau$}}}\in [0,1]}\sum_{\bm y,\bm x_{\scalebox{.8}{\tiny$\tau^c$}}} \sum_{\bm u_{\scalebox{.8}{\tiny$\tau^c$}}\in \Au[\tau^c]{i_{\tau^c}}}
P^n_{\bm U_{\tau^c}}(\bm u_{\tau^c})\Qu[\tau^c]{i_{\tau^c} }^n(\bm x_{\tau^c})
\hspace{10.5em}\nonumber \\
 G_{i_\tau}(1-\Sf[i_\tau]{}{j_\tau}\ro[i_\tau]{}{j_\tau},\bm x_{\tau^c},\bm y)G_{j_\tau}(\Sf[i_\tau]{}{j_\tau},\bm x_{\tau^c},\bm y)^{\ro[i_\tau]{}{j_\tau}}
.\label{itw5}
 \end{align}
 
Applying  H\"{o}lder's inequality in the form of
 \begin{equation}
 	\sum_i C_ia_ib_i\leq \bigg(\sum_iC_ia_i^{\frac{1}{p}}\bigg)^p \bigg(\sum_iC_ia_i^{\frac{1}{1-p}}\bigg)^{1-p}
 \end{equation}
for $p\in[0,1]$, into \eqref{itw5}, we obtain
 \begin{align}
 \bar{\epsilon}_\tau^n\leq 
\sum_{j_{\scalebox{.8}{\tiny$\tau$}},i_{\scalebox{.8}{\tiny$\tau$}}=1,2}
\min_{\ro[i_\tau]{}{j_\tau}\in [0,1]}  F_{i_\tau}^n\biggl(1-\Sf[i_\tau]{}{j_\tau}\ro[i_\tau]{}{j_\tau},\frac{1}{\Pu[i_\tau]{}{j_\tau}}\biggr)^{\Pu[i_\tau]{}{j_\tau}}
F_{j_\tau}^n\biggl(\Sf[i_\tau]{}{j_\tau},\frac{\ro[i_\tau]{}{j_\tau}}{1-\Pu[i_\tau]{}{j_\tau}}\biggr)^{1-\Pu[i_\tau]{}{j_\tau}}
,\label{itw6}
 \end{align}
 where
 \begin{align}
 F_{j_\tau}^n(a,b)=
\sum_{i_{\tau^c}=1,2} \sum_{\bm u_{\scalebox{.8}{\tiny$\tau^c$}}\in \Au[\tau^c]{i_{\tau^c}} }
\sum_{\bm x_{\scalebox{.8}{\tiny$\tau^c$}},\bm y}
P^n_{\bm U_{\tau^c}}(\bm u_{\tau^c})
 \Qu[\tau^c]{i_{\tau^c} }^n(\bm x_{\tau^c})G_{j_\tau}(a,\bm x_{\tau^c},\bm y)^{b}.\label{F_def}
 \end{align}
 Now, by setting $\Sf[i_\tau]{}{j_\tau}=\frac{1}{1+\ro[j_\tau]{}{}}$, $\ro[i_\tau]{}{j_\tau}=\frac{\ro[i_\tau]{}{}(1+\ro[j_\tau]{}{})}{1+\ro[i_\tau]{}{}}$ and $\Pu[i_\tau]{}{j_\tau}=\frac{1}{1+\ro[i_\tau]{}{}}$, the average error probability can be written  as 
  \begin{align}
 \bar{\epsilon}_\tau^n\leq 
\sum_{j_{\scalebox{.8}{\tiny$\tau$}},i_{\scalebox{.8}{\tiny$\tau$}}=1,2}
\min_{\ro[i_\tau]{}{}, \ro[j_\tau]{}{}\in[0,1]}
F_{i_\tau}^n\biggl(\frac{1}{1+\ro[i_\tau]{}{}},1+\ro[i_\tau]{}{}\biggr)^{\frac{1}{1+\ro[i_\tau]{}{}}}
 F_{j_\tau}^n\biggl(\frac{1}{1+\ro[j_\tau]{}{}},1+\ro[j_\tau]{}{}\biggr)^{\frac{\ro[i_\tau]{}{}}{1+\ro[i_\tau]{}{}}}.\label{itw7}
 \end{align}

 Since $F_{i_\tau}^n(\cdot), F_{j_\tau}^n(\cdot)\geq 0$ and $\frac{1}{1+\ro[i_\tau]{}{}}+\frac{\ro[i_\tau]{}{}}{1+\ro[i_\tau]{}{}}=1$, by using weighted arithmetic-geometric inequality, \eqref{itw7} is bounded as
   \begin{align}
 \bar{\epsilon}_\tau^n\leq 
\sum_{j_{\scalebox{.8}{\tiny$\tau$}},i_{\scalebox{.8}{\tiny$\tau$}}=1}^2
\min_{\ro[i_\tau]{}{}, \ro[j_\tau]{}{}\in[0,1]}
\frac{1}{1+\ro[i_\tau]{}{}} F_{i_\tau}^n\biggl(\frac{1}{1+\ro[i_\tau]{}{}},1+\ro[i_\tau]{}{}\biggr)
+\frac{\ro[i_\tau]{}{}}{1+\ro[i_\tau]{}{}}F_{j_\tau}^n\biggl(\frac{1}{1+\ro[j_\tau]{}{}},1+\ro[j_\tau]{}{}\biggr),
\label{itw8}
\end{align}
where by rearranging the terms of the sum, we have
\begin{align}
 \bar{\epsilon}_\tau^n\leq \sum_{i_{\scalebox{.8}{\tiny$\tau$}}=1,2}
\min_{\ro[i_\tau]{}{}, \ro[j_\tau]{}{}\in[0,1]}
F_{i_\tau}^n\biggl(\frac{1}{1+\ro[i_\tau]{}{}},1+
\ro[i_\tau]{}{}
\biggr)
 \sum_{j_{\scalebox{.8}{\tiny$\tau$}}=1,2}
\biggl(\frac{1}{1+\ro[i_\tau]{}{}}+\frac{\ro[j_\tau]{}{}}{1+\ro[j_\tau]{}{}}\biggr).\label{itw9}
 \end{align}


Next, we may use the following Lemma.
\begin{lemma}
\label{itw_lema_1}
For a given $\rho\in[0,1]$, and $F_{j_\tau}^n(a,b)$ defined in \eqref{F_def}, the following inequality holds
\begin{align}
-\frac{1}{n}\log\Biggl(F_{j_\tau}^n\biggl(\frac{1}{1+\rho},1+\rho\biggr)\Biggr)\geq \min_{i_{\tau^c}=1,2} E_0(\rho,\Qu[\tau]{j_\tau},W\Qu[\tau^c]{i_{\tau^c}})\hspace{10em}\nonumber\\
-\Es[j_\tau]{}(\rho,P_{ U_\tau},\gamma_\tau)-\Es[i_{\tau^c}]{}(0,P_{ U_{\tau^c}},\gamma_{\tau^c})-\frac{1}{n}\log(2), \label{pr2_lem1_itw}
\end{align}
where
$E_0(\cdot)$ is given by \eqref{E0_func} and $\Es[i]{}(\cdot)$ for $i=1,2$  is given by \eqref{Esi_itw_2018_1} and \eqref{Esi_itw_2018_2}.
\end{lemma}
\begin{IEEEproof}
\label{proof_itw_lema_1}
In order to prove Lemma \ref{itw_lema_1}, we recall that by inserting  $G_{j_\tau}\left(\frac{1}{1+\rho},\bm x_{\tau^c},\bm y\right)$ defined in \eqref{G_def} into \eqref{F_def}, $F_{j_\tau}^n\left(\frac{1}{1+\rho},1+\rho\right)$ can be written as
 \begin{align}
F_{j_\tau}^n\biggl(\frac{1}{1+\rho},1+\rho\biggr)=
\sum_{i_{\tau^c}=1,2}
\sum_{\bm x_{\scalebox{.8}{\tiny$\tau^c$}},\bm y}\Qu[\tau^c]{i_{\tau^c} }^n(\bm x_{\tau^c})\Biggl(
\sum_{\bm  x_\tau}
\Qu[\tau]{i_\tau }^n(\bm{x}_\tau)
W^n(\bm {y}|\bm x_{\scalebox{0.8}{\tiny$\tau$}},\bm x_{\scalebox{0.8}{\tiny$\tau^c$}})^{\frac{1}{1+\rho}} \Biggr)^{1+\rho}
\hspace{3em}\nonumber\\ 
\times\sum_{\bm u_{\scalebox{.8}{\tiny$\tau^c$}}\in \Au[\tau^c]{i_{\tau^c}} }P^n_{\bm U_{\tau^c}}(\bm u_{\tau^c})
\Biggl(\sum_{\bm u_{\scalebox{.8}{\tiny$\tau$}}
 \in \Au[\tau]{i_\tau}
}P^n_{\scalebox{1}{\tiny$\bm U$}_{\scalebox{.8}{\tiny$\tau$}}}(\bm{u}_{\scalebox{0.8}{\tiny$\tau$}})^{\frac{1}{1+\rho}}\Biggr)^{1+\rho}.
\label{F_rho_simple_2018}
 \end{align}
Applying the  identity $\sum_{u\in A}f(u)=\sum_u f(u)\mathbbm{1}\{u\in A\}$ to the summation over $\bm u_{\nu}\in \Au[\nu]{i_{\nu}}, {\nu=\tau,\tau^{c}}$  of \eqref{F_rho_simple_2018}, 
we obtain
\begin{align}
F_{j_\tau}^n\biggl(\frac{1}{1+\rho},1+\rho\biggr)=\sum_{i_{\tau^c}=1,2} e^{-E_0\big(\rho,\Qu[\tau]{j_\tau}^n,W^n\Qu[\tau]{i_{\tau^c}}\big)}
\sum_{\bm u_{\scalebox{.8}{\tiny$\tau^c$}}}P^n_{\bm U_{\tau^c}}(\bm u_{\tau^c})\mathbbm{1}\Bigl\{\bm u_{\scalebox{.8}{\tiny$\tau^c$}}\in\Au[\tau^c]{i_{\tau^c}}\Bigr\}
\hspace{3em}
\nonumber\\
\times\Biggl(\sum_{\bm u_{\scalebox{.8}{\tiny$\tau$}}
}P^n_{\scalebox{1}{\tiny$\bm U$}_{\scalebox{.8}{\tiny$\tau$}}}(\bm{u}_{\scalebox{0.8}{\tiny$\tau$}})^{\frac{1}{1+\rho}}\mathbbm{1}\Bigl\{\bm u_{\scalebox{.8}{\tiny$\tau$}}
 \in \Au[\tau]{i_\tau}\Bigr\}\Biggr)^{1+\rho},
 \hspace{.5em}
 \label{F_16.4.2018}
\end{align}
where in \eqref{F_16.4.2018}, in view of \eqref{E0_func} we applied $\sum_{b}f_b \cdot \big(\sum_a g_a\big)^c=\sum_b\big(\sum_a g_a \cdot f_b^{1/c}\big)^c$ into the first summation of \eqref{F_rho_simple_2018} and we expressed it in terms of $E_0$ function.

Next, we focus on the summations over $\bm u_\tau$ and $\bm u_{\tau^c}$ in \eqref{F_16.4.2018}. 
Let $\nu=\tau,\tau^c$, in view of \eqref{A1_itw2018} and \eqref{A2_itw2018}, for a given $\bm u_{\nu}$, we have
$\mathbbm{1}\bigl\{\bm u_{\nu}
 \in \Au[\nu]{1}\bigl\}=\mathbbm{1}\bigl\{P_{\bm U_\nu}^n(\bm u_\nu)\geq \gamma_\nu^n\}$ and 
 $\mathbbm{1}\bigl\{\bm u_{\nu}
 \in \Au[\nu]{i_2}\bigl\}=\mathbbm{1}\bigl\{P_{\bm U_\nu}^n(\bm u_\nu)< \gamma_\nu^n\}$. Considering this fact and applying 
 $\mathbbm{1}\bigl\{a\leq b\bigl\}\leq \big(\frac{b}{a}\big)^\lambda$ for $\lambda\geq 0$  to all indicator functions of \eqref{F_16.4.2018}, we find that
 \begin{align}
F_{j_\tau}^n\biggl(\frac{1}{1+\rho},1+\rho\biggr)\leq \min_{\lambda_{\tau}, \lambda_{\tau^c}\geq 0}\sum_{i_{\tau^c}=1,2} e^{-E_0\big(\rho,\Qu[\tau]{j_\tau}^n,W^n\Qu[\tau]{i_{\tau^c}}\big)}
\hspace{12em}
\nonumber\\
\times\sum_{\bm u_{\scalebox{.8}{\tiny$\tau^c$}}}P^n_{\bm U_{\tau^c}}(\bm u_{\tau^c})\biggl(\frac{\gamma^n_{\tau^c}}{P^n_{\bm U_{\tau^c}}(\bm u_{\tau^c})}\biggr)^{(-1)^{i_{\tau^c}}\lambda_{\tau^c}}
\Biggl(\sum_{\bm u_{\scalebox{.8}{\tiny$\tau$}}
}P^n_{\scalebox{1}{\tiny$\bm U$}_{\scalebox{.8}{\tiny$\tau$}}}(\bm{u}_{\scalebox{0.8}{\tiny$\tau$}})^{\frac{1}{1+\rho}}
\biggl(\frac{\gamma^n_{\tau}}{P^n_{\bm U_{\tau}}(\bm u_{\tau})}\biggr)^{\frac{(-1)^{i_{\tau}}\lambda_{\tau}}{1+\rho}}\Biggr)^{1+\rho},
 \label{F_16.4.2018_12.50}
\end{align}
where in \eqref{F_16.4.2018_12.50} we tightened the bound by minimizing the objective function over $\lambda_{\tau}, \lambda_{\tau^c}\geq 0$.

Using Lemma \ref{itw_lemma_2} in Appendix \ref{lema_util_itw}, the second and the third terms of 
\eqref{F_16.4.2018_12.50} can be expressed in terms of the $E_{si}(\cdot)$ function at $\rho=0$ and arbitrary $\rho$, respectively. Doing so, we obtain that
\begin{align}
F_{j_\tau}^n\biggl(\frac{1}{1+\rho},1+\rho\biggr)\leq \sum_{i_{\tau^c}=1,2}
  e^{-E_0\big(\rho,\Qu[\tau]{j_\tau}^n,W^n\Qu[\tau]{i_{\tau^c}}\big)}
 \times e^{\Es[j_\tau]{}\big(\rho,P_{\bm U_\tau}^n,\gamma_\tau^n\big)+\Es[i_{\tau^c}]{}\big(0,P_{\bm U_{\tau^c}}^n,\gamma_{\tau^c}^n\big)}.\label{pr_lem1_itw}
\end{align}
Finally, we bound each term in the summation in  \eqref{pr_lem1_itw} by the maximum term, use that the sources and the channel are memoryless, and taking logarithms, we obtain to \eqref{pr2_lem1_itw}.
\end{IEEEproof}

Next, upper bounding \eqref{itw9} by 
 the maximum term over $i_{\scalebox{.8}{\tiny$\tau$}}$, further upper bounding by the worst type of error $\tau$, taking logarithms and using \eqref{pr2_lem1_itw}, after some mathematical manipulations we find that the exponential decay of $\bar \epsilon^n$ is given by
\begin{align}
-\frac{1}{n}\log (\bar \epsilon^n) \geq \min_\tau \min_{i_{\scalebox{.8}{\tiny$\tau$}},i_{\scalebox{.8}{\tiny$\tau^c$}}} \max_{\rho\in[0,1]}
E_0\big(\rho,\Qu[\tau]{i_\tau},W\Qu[\tau^c]{i_{\tau^c}}\big)\hspace{15em}\nonumber\\
-\Es[i_\tau]{}\big(\rho,P_{ U_\tau},\gamma_\tau\big)-\Es[i_{\tau^c}]{}\big(0,P_{ U_{\tau^c}},\gamma_{\tau^c}\big)-\frac{\log(o(n))}{n},
 \end{align}
 where $o(n)$ is a sequence satisfying $\lim_{n\rightarrow\infty}\frac{o(n)}{n}=0$. Using the following properties
\begin{gather}
	\liminf_{n\rightarrow \infty} (a_n+b_n)\geq \liminf_{n\rightarrow \infty} a_n+\liminf_{n\rightarrow \infty} b_n\\
	\liminf_{n\rightarrow \infty}\min \{a_n,b_n\}= \min \big\{\liminf_{n\rightarrow \infty} a_n,\liminf_{n\rightarrow \infty} b_n\big\},\\
		\liminf_{n\rightarrow \infty}\max \{a_n\}\geq \max \big\{\liminf_{n\rightarrow \infty} a_n\big\},
\end{gather}
we further obtain that
\begin{align}
\liminf_{n\rightarrow\infty} -\frac{1}{n}\log (\bar \epsilon^n) \geq \min_{\tau} \min_{i_{\scalebox{.8}{\tiny$\tau$}},i_{\scalebox{.8}{\tiny$\tau^c$}}} \max_{\rho\in[0,1]}
E_0\big(\rho,\Qu[\tau]{i_\tau},W\Qu[\tau^c]{i_{\tau^c}}\big)\hspace{12em}\nonumber\\
-\Es[i_\tau]{}\big(\rho,P_{ U_\tau},\gamma_\tau\big)-\Es[i_{\tau^c}]{}\big(0,P_{ U_{\tau^c}},\gamma_{\tau^c}\big).\label{last_term_itw}
\end{align}
Finally,  we optimize \eqref{last_term_itw}
over $\gamma_\nu$ for $\nu=1,2$. This concludes the proof.

\section{Proof of Proposition \ref{itw_prop1}}
\label{proof_prop2_itw}
Now, we focus on $F_{\tau,i_{\scalebox{.8}{\tiny$\tau$}},i_{\scalebox{.8}{\tiny$\tau^c$}}}(\munderbar \gamma)$ given in \eqref{eq:F}. Let $i_1=1$ for an arbitrary  $\tau$. Since $\gamma_1$ and $\gamma_2$ are independent from each other, regardless the value of $i_2$, the function
$F_{\tau,i_{\scalebox{.8}{\tiny$\tau$}},i_{\scalebox{.8}{\tiny$\tau^c$}}}(\munderbar \gamma)$ is of the form $\max_{\rho} E(\rho)-\Es[1]{}(\rho,P_{U_1},\gamma_1)$. Then, using Lemma \ref{itw_lema_5}, we have that $F_{\tau,i_{\scalebox{.8}{\tiny$\tau$}},i_{\scalebox{.8}{\tiny$\tau^c$}}}(\munderbar \gamma)$ is non-decreasing with respect to $\gamma_1$. Similarly, when $i_1=2$, we have that $F_{\tau,i_{\scalebox{.8}{\tiny$\tau$}},i_{\scalebox{.8}{\tiny$\tau^c$}}}(\munderbar \gamma)$ is of the form $\max_{\rho} E(\rho)-\Es[2]{}(\rho,P_{U_1},\gamma_1)$ so that it
is non-increasing with respect to $\gamma_1$. The same reasoning applies for $i_2$. That is,
$F_{\tau,i_{\scalebox{.8}{\tiny$\tau$}},i_{\scalebox{.8}{\tiny$\tau^c$}}}(\munderbar \gamma)$ is non-decreasing with respect to $\gamma_2$ for $i_2=1$, and non-increasing with respect to $\gamma_2$ for $i_2=2$, always regardless of the value of $i_1$. 
 
Using the fact that the minimum of  monotonic functions is monotonic, we conclude that $f_{i_1,i_2}(\munderbar \gamma)$ given in \eqref{nef_al_itw_27_4} is non-decreasing with respect to $\gamma_1$ when $i_1=1$, and non-increasing with respect to $\gamma_1$ when $i_1=2$.  Similarly, $f_{i_1,i_2}(\munderbar \gamma)$ is non-decreasing (non-increasing) with respect to $\gamma_2$ when $i_2=1$ ($i_2=2$).

Writing equation \eqref{itw_13_32_4_19} as
\begin{equation}
	\max_{\gamma_1} \max_{\gamma_2} \min_{i_2} \min_{i_1} f_{i_1,i_2}(\gamma_1,\gamma_2),
\end{equation}
we note that, for a fixed $\gamma_1$, the optimization problem $\max_{\gamma_2} \min_{i_2}  \min_{i_1} f_{i_1,i_2}(\gamma_1,\gamma_2)$ satisfies Lemma \ref{itw_lema4} with $\gamma=\gamma_2$, $i=i_2$, and $k_i(\gamma)=\min_{i_1} f_{i_1,i}(\gamma_1,\gamma)$.  Therefore, the optimal $\gamma_2^\star$ satisfies 
 \begin{align}
\min_{i_1=1,2} f_{i_1,1}(\gamma_1, \gamma_2^\star)=\min_{i_1=1,2} f_{i_1,2}(\gamma_1, \gamma_2^\star),\label{itw_14_42_4_19}
\end{align}
 whenever \eqref{itw_14_42_4_19} has solution. Otherwise, we have $\gamma_2^\star=0$ when $f_{i_1,1}(\gamma_1, 0)> f_{i_1,2}(\gamma_1,0)$, or $\gamma_2^\star=1$ when $f_{i_1,1}(\gamma_1, 0)\leq  f_{i_1,2}(\gamma_1,0)$.
 

Setting $\gamma_2=\gamma_2^\star$, the optimization problem $\max_{\gamma_1} \min_{i_1}  \min_{i_2} f_{i_1,i_2}(\gamma_1,\gamma_2^\star)$ satisfies Lemma \ref{itw_lema4} with $\gamma=\gamma_1$, $i=i_1$, and $k_i(\gamma)=\min_{i_2} f_{i,i_2}(\gamma,\gamma_2^\star)$. Hence, $\gamma_1^\star$ maximizing \eqref{itw_13_32_4_19} satisfies
\begin{align}
\min_{i_2=1,2} f_{1,i_2}(\gamma_1^\star,\gamma_2^\star)=\min_{i_2=1,2} f_{2,i_2}(\gamma_1^\star,\gamma_2^\star),\label{itw_14_48_4_19}
\end{align}
and in the case \eqref{itw_14_48_4_19} does not have solution,
$\gamma_1^\star=0$ when $f_{1,i_2}(0,\gamma_2)> f_{2,i_2}(0,\gamma_2)$,  or $\gamma_1^\star=1$ otherwise. Combining \eqref{itw_14_42_4_19} and \eqref{itw_14_48_4_19} we obtain \eqref{utw_optimal_gamma}.

\section{Proof of the Upper Bound for the Achievable Exponent}
\label{proof_pre_up_bound_itw}
In view of the  max-min inequality \cite{fans_minimax},
after upper bounding 
\eqref{our_exp_itw} by
 swapping the maximization over $\gamma_1$,$\gamma_2$ 
 with the minimization over $\tau$,
the upper bound  given by \eqref{upb_itw},
 follows immediately from the
following Lemma.
 \begin{lemma}
 \label{itw_lema3}
For a given $\tau=\{\{1\},\{2\},\{1,2\}\}$, we have
\begin{align}
\max_{\gamma_1,\gamma_2\in[0,1]}\min_{i_{\scalebox{.8}{\tiny$\tau$}},i_{\scalebox{.8}{\tiny$\tau^c$}}=1,2} \max_{\rho\in[0,1]}
E_0(\rho,\Qu[\tau]{i_\tau},W\Qu[\tau^c]{i_{\tau^c}})
-\Es[i_\tau]{}(\rho,P_{ U_\tau},\gamma_\tau)-\Es[i_{\tau^c}]{}(0,P_{ U_{\tau^c}},\gamma_{\tau^c})
\hspace{2em}\nonumber\\
\leq\max_{i_{\tau^c}=1,2}
\max_{\rho\in[0,1]} \bar E_0(\rho, \{\Qu[\tau]{1},\Qu[\tau]{2}\},W\Qu[\tau^c]{i_{\tau^c}})-E_s(\rho,P_{ U_\tau}),\label{formula_lema3_itw}
\end{align}
where equality holds for $\tau=\{\{1\},\{2\}\}$.
 \end{lemma}
 \begin{IEEEproof}
 Firstly, we consider $\tau=\{\{1\},\{2\}\}$. In this case, by
focusing on the optimization problem given on the left hand side of \eqref{formula_lema3_itw}, we may note that since $\Es[i_{\tau^c}]{}(0,P_{ U_{\tau^c}},\gamma_{\tau^c})$ does not depend on $\rho$,
 the maximization over $\rho$ of the left hand side of  \eqref{formula_lema3_itw} is done independently from $\Es[i_{\tau^c}]{}(0,P_{ U_{\tau^c}},\gamma_{\tau^c})$. Additionally,
in view of \eqref{esi_for_gamma0} and  \eqref{esi_for_gamma1}, we may note that 
by moving $\gamma_{\tau^c}$ along the $[0,1]$ interval,
 $\Es[1]{}(0,P_{ U_{\tau^c}},\gamma_{\tau^c})$ decreases from zero to $-\infty$, while $\Es[2]{}(0,P_{ U_{\tau^c}},\gamma_{\tau^c})$ increases from $-\infty$ to zero.
Hence, the minimum over $i_\tau$ and $i_{\tau^c}$ of  
\begin{align}
\max_{\rho\in[0,1]}
E_0(\rho,\Qu[\tau]{i_\tau},W\Qu[\tau^c]{i_{\tau^c}})
-\Es[i_\tau]{}(\rho,P_{ U_\tau},\gamma_\tau)
-\Es[i_{\tau^c}]{}(0,P_{ U_{\tau^c}},\gamma_{\tau^c}),\label{itw_12_21_4_18}
\end{align}
is attained at  $\gamma_{\tau^c}=0$ for $i_{\tau^c}=1$, or $\gamma_{\tau^c}=1$ for $i_{\tau^c}=2$, both leading to $\Es[i_{\tau^c}]{}(0,P_{ U_{\tau^c}},\gamma_{\tau^c})=0$. As a result, it is sufficient to consider $\max_{\gamma_{\tau^c}\in\{0,1\}}$ instead of $\max_{\gamma_{\tau^c}\in[0,1]}$. This implies that the left hand side of \eqref{formula_lema3_itw} can be written as
\begin{align}
	\max \bigg\{
	\max_{\gamma_\tau\in[0,1]} \min_{i_\tau} \max_{\rho\in[0,1]}
E_0(\rho,\Qu[\tau]{i_\tau},W\Qu[\tau^c]{i_{\tau^c}})
-\Es[i_\tau]{}(\rho,P_{ U_\tau},\gamma_\tau) \bigg|_{i_{\tau^c}=1, \gamma_{\tau^c}=0}	, \nonumber \hspace{5em} \\
	\max_{\gamma_\tau\in[0,1]} \min_{i_\tau} \max_{\rho\in[0,1]}
E_0(\rho,\Qu[\tau]{i_\tau},W\Qu[\tau^c]{i_{\tau^c}})
-\Es[i_\tau]{}(\rho,P_{ U_\tau},\gamma_\tau) \bigg|_{i_{\tau^c}=2, \gamma_{\tau^c}=1}	
	\bigg\},
\end{align}
or equivalently 
 \begin{align}
 \max_{i_{\scalebox{.8}{\tiny$\tau^c$}}=1,2}\max_{\gamma_\tau\in[0,1]}\min_{i_{\scalebox{.8}{\tiny$\tau$}}=1,2} \max_{\rho\in[0,1]}
E_0(\rho,\Qu[\tau]{i_\tau},W\Qu[\tau^c]{i_{\tau^c}})
-\Es[i_\tau]{}(\rho,P_{ U_\tau},\gamma_\tau).\label{itw_14_33_18_4}
 \end{align}
Equation \eqref{itw_14_33_18_4} can be interpreted 
as an achievable exponent for a point-to-point channel with transition-probability $W\Qu[\tau^c]{i_{\tau^c}}$, a pair of distributions $\{\Qu[\tau]{1},\Qu[\tau]{2} \}$ and a partition of the source message set into two classes.  This problem  is well-studied in \cite{jscc}. In fact, $i_{\tau^c}$ in  \eqref{itw_14_33_18_4} is just a parameter selecting either
$W\Qu[\tau^c]{1}$ or $W\Qu[\tau^c]{2}$. From  \cite[Theorem 2]{jscc}, equation \eqref{itw_14_33_18_4} is equal to
 \begin{align}
 \max_{i_{\scalebox{.8}{\tiny$\tau^c$}}=1,2} \max_{\rho\in[0,1]} \bar E_0(\rho, \{\Qu[\tau]{1},\Qu[\tau]{2}\},W\Qu[\tau^c]{i_{\tau^c}})-E_s(\rho,P_{ U_\tau}),
 \end{align}
 which leads \eqref{formula_lema3_itw} for type $\tau\in\{\{1\},\{2\}\}$.

For $\tau=\{1,2\}$, in view of the min-max inequality \cite{fans_minimax}, we upper bound the left hand side of
 \eqref{formula_lema3_itw}
by swapping the maximization over $\gamma_2$ with the minimization over $i_1$ as
 \begin{align}
\max_{\gamma_1\in[0,1]}
\min\Bigl\{T_1(\gamma_1), T_2(\gamma_1)\Bigr\},
\label{formula_lema3_itw_27_4_2018_20_26}
\end{align}
where
 \begin{align}
T_1(\gamma_1)=\max_{\gamma_2\in[0,1]}
\min_{i_{2}=1,2} \max_{\rho\in[0,1]}
E_0(\rho,\Qu[1]{1}\Qu[2]{i_{2}},W)
-\Es[1]{}(\rho,P_{ U_1},\gamma_1)
-\Es[i_{2}]{}(\rho,P_{ U_{2}},\gamma_2),
\label{T1_itw_27_4_2018_20_26}
\end{align}
and
 \begin{align}
T_2(\gamma_1)=\max_{\gamma_2\in[0,1]}
\min_{i_{2}=1,2} \max_{\rho\in[0,1]}
E_0(\rho,\Qu[1]{2}\Qu[2]{i_{2}},W)
-\Es[2]{}(\rho,P_{ U_1},\gamma_1)
-\Es[i_{2}]{}(\rho,P_{ U_{2}},\gamma_2).
\label{T2_itw_27_4_2018_20_26}
\end{align}
We note that $\Es[1]{}(\rho,P_{ U_1},\gamma_1)$ in \eqref{T1_itw_27_4_2018_20_26} does not change with $i_2$ and $\gamma_2$.  
Thus, the optimization problem  \eqref{T1_itw_27_4_2018_20_26} can be seen as a refined  achievable exponent for a point-to-point channel with a new $E_0$ function as $E_0(\rho,\Qu[1]{1}\Qu[2]{i_{2}},W)
-\Es[1]{}(\rho,P_{ U_1},\gamma_1)$ having two input distributions $ \{\Qu[1]{1}\Qu[2]{1},\Qu[1]{1}\Qu[2]{2}\}$, and a partition of a source message into two classes.  Equation 
 \eqref{T1_itw_27_4_2018_20_26} can be 
  written  in terms of 
  the concave hull of $\max_{i_2\in\{1,2\}} E_0(\rho,\Qu[1]{1}\Qu[2]{i_{2}},W)
-\Es[1]{}(\rho,P_{ U_1},\gamma_1)$. Since $\Es[1]{}(\rho,P_{ U_1},\gamma_1)$ is a convex function with respect to $\rho$, using Lemma \ref{itw_lema_concavehull}  we upper bound the concave hull of $\displaystyle\max_{i_2\in\{1,2\}} E_0(\rho,\Qu[1]{1}\Qu[2]{i_{2}},W)
-\Es[1]{}(\rho,P_{ U_1},\gamma_1)$ by $
 \bar E_0(\rho, \{\Qu[1]{1}\Qu[2]{1},\Qu[1]{1}\Qu[2]{2}\},W)
-\Es[1]{}(\rho,P_{ U_1},\gamma_1)$. Therefore, from applying \cite[Theorem 2]{jscc}, $T_1(\gamma_1)$ is upper bounded as
\begin{align}
T_1(\gamma_1)\leq \max_{\rho\in[0,1]} \bar E_0(\rho, \{\Qu[1]{1}\Qu[2]{1},\Qu[1]{1}\Qu[2]{2}\},W)
-\Es[1]{}(\rho,P_{ U_1},\gamma_1) -E_s(\rho,P_{ U_2}).\label{itw_bem1}
\end{align}
Similarly,
  \begin{align}
 T_2(\gamma_1)\leq \max_{\rho\in[0,1]} \bar E_0(\rho, \{\Qu[1]{2}\Qu[2]{1},\Qu[1]{2}\Qu[2]{2}\},W)
-\Es[2]{}(\rho,P_{ U_1},\gamma_1) -E_s(\rho,P_{ U_2}).\label{itw_bem2}
  \end{align}  
 Inserting the right hand sides of \eqref{itw_bem1} and \eqref{itw_bem2}
into \eqref{formula_lema3_itw_27_4_2018_20_26}, we obtain
\begin{align}
\max_{\gamma_1\in[0,1]}
\min\Bigl\{T_1(\gamma_1), T_2(\gamma_1)\Bigr\} \leq 
\max_{\gamma_1}
\min_{i_1\in\{1,2\}}
\max_{\rho\in[0,1]} \bar E_0(\rho, \{\Qu[1]{i_1}\Qu[2]{1},\Qu[1]{i_1}\Qu[2]{2}\},W) \nonumber \hspace{2em} \\
-\Es[i_1]{}(\rho,P_{ U_1},\gamma_1)
-E_s(\rho,P_{ U_2}). \label{itw_29_4_2018_12_34}
\end{align}

Again, the right hand side of \eqref{itw_29_4_2018_12_34} can be written in terms of the concave hull of  the function $\bar E_0(\rho, \{\Qu[1]{i_1}\Qu[2]{1},\Qu[1]{i_1}\Qu[2]{2}\},W)
-E_s(\rho,P_{ U_2})$. Since $E_s(\rho,P_{ U_2})$ is convex in $\rho$, we apply Lemma \ref{itw_lema_concavehull} again to upper bound the concave hull of $\bar E_0(\rho, \{\Qu[1]{i_1}\Qu[2]{1},\Qu[1]{i_1}\Qu[2]{2}\},W)
-E_s(\rho,P_{ U_2})$ by $\bar E_0(\rho, \{\Qu[1]{1},\Qu[1]{2},\Qu[2]{1},\Qu[2]{2}\},W)
-E_s(\rho,P_{ U_2})$. Finally using \cite[Theorem 2]{jscc}, we obtain that \eqref{formula_lema3_itw_27_4_2018_20_26}
is upper bounded by 
\begin{align}
\max_{\rho} \bar E_0(\rho, \{\Qu[1]{1},\Qu[1]{2},\Qu[2]{1},\Qu[2]{2}\},W)
 -E_s(\rho,P_{ U_1})-E_s(\rho,P_{ U_2}).\label{itw_29_4_2018_13_13}
\end{align}
\end{IEEEproof}


\section{}
\label{lema_util_itw}
In this appendix, we provide a number of general equations and lemmas that will be used through the
paper. Throughout this Appendix, we consider a discrete memoryless source characterized by a probability distribution $P_U$.

\begin{lemma}
\label{itw_lemma_2}
Let $i=1,2$, for a given source probability distribution $P_U$ and some $\gamma\in[0,1]$. Then, we have that
\begin{align}
\min_{\lambda\geq 0}
\Biggl(\sum_{\bm u
}P^n_{\scalebox{1}{\tiny$\bm U$}}(\bm{u})^{\frac{1}{1+\rho}}
\biggl(\frac{\gamma^n}{P^n_{\bm U}(\bm u)}\biggr)^{\frac{(-1)^{i}\lambda}{1+\rho}} \Biggr)^{1+\rho}= e^{\Es[i]{}(\rho,P_{\scalebox{.8}{\tiny$\bm U$}}^n,\gamma^n)},\label{formula_itw_lemma2}
\end{align}
where $\Es[i]{}(\rho,P_U,\gamma)$ for $i=1,2$ is given by \eqref{Esi_itw_2018_1} and  \eqref{Esi_itw_2018_2}.
\end{lemma}

\begin{IEEEproof}
In order to prove \eqref{formula_itw_lemma2}, we may note that since the objective function in  \eqref{formula_itw_lemma2} is convex with respect to $\lambda$, the optimal $\lambda^\star$ satisfies
\begin{align}
\left.\frac{\partial}{\partial \lambda} \left(\sum_{\bm u
}P^n_{\scalebox{1}{\tiny$\bm U$}}(\bm{u})^{\frac{1}{1+\rho}}
\Bigl(\frac{\gamma^n}{P^n_{\bm U}(\bm u)}\Bigr)^{\frac{(-1)^{i}\lambda}{1+\rho}}\right)^{1+\rho}\right|_{\lambda^\star\geq 0}=0.
\end{align}
This leads to
\begin{align}
\frac{\sum_{\bm u
}P^n_{\scalebox{1}{\tiny$\bm U$}}(\bm{u})^{\frac{1-(-1)^{i}\lambda^\star}{1+\rho}}
\log(P^n_{\scalebox{1}{\tiny$\bm U$}}(\bm{u}))}{\sum_{\bm u
}P^n_{\scalebox{1}{\tiny$\bm U$}}(\bm{u})^{\frac{{1-(-1)^{i}\lambda^\star}}{1+\rho}}
}=\log(\gamma^n).\label{itw_kh1}
\end{align}
It is convenient to  define $\rho_\gamma$ through the implicit equation
\begin{equation}
	\frac{1-(-1)^{i}\lambda^\star}{1+\rho}=\frac{1}{1+\rho_\gamma}.
	\label{relation}
\end{equation}
When the solution to \eqref{itw_kh1} is strictly negative, i.e., when
\begin{equation}
	(-1)^{i}\biggl(\frac{1}{1+\rho} -\frac{1}{1+\rho_\gamma}\biggr)< 0,
	\label{negativelambda}
\end{equation}
we have  $\lambda^\star=0$, and hence  \eqref{formula_itw_lemma2} simplifies to
\begin{align}
\Bigg(\sum_{\bm u
}P^n_{\scalebox{1}{\tiny$\bm U$}}(\bm{u})^{\frac{1}{1+\rho}}
\biggl(\frac{\gamma^n}{P^n_{\bm U}(\bm u)}\biggr)^{\frac{(-1)^{i}\lambda}{1+\rho}}\Bigg)^{1+\rho}\Bigg|_{\lambda=0}=
\bigg(\sum_{\bm u
}P^n_{\scalebox{1}{\tiny$\bm U$}}(\bm{u})^{\frac{1}{1+\rho}}
\bigg)^{1+\rho}
=e^{E_s(\rho,P^n_{\scalebox{.8}{\tiny$\bm U$}}(\bm{u}))}.\label{itw_kh1.5}
\end{align}
Otherwise, when the solution to \eqref{itw_kh1} is non-negative, i.e., when
\begin{equation}
	(-1)^{i}\biggl(\frac{1}{1+\rho} -\frac{1}{1+\rho_\gamma}\biggr)\geq 0
	\label{positivelambda}
\end{equation}
and using \eqref{relation}, the left hand side of \eqref{formula_itw_lemma2} satisfies
\begin{align}
\min_{\lambda\geq 0}
\Bigg(\sum_{\bm u
}P^n_{\scalebox{1}{\tiny$\bm U$}}(\bm{u})^{\frac{1}{1+\rho}}
\biggl(\frac{\gamma^n}{P^n_{\bm U}(\bm u)}\biggr)^{\frac{(-1)^{i}\lambda}{1+\rho}}\Bigg)^{1+\rho}
\hspace{19em}\nonumber\\
=\bigg(\sum_{\bm u
}P^n_{\scalebox{1}{\tiny$\bm U$}}(\bm{u})^{\frac{1}{1+\rho_\gamma}} \bigg)^{1+\rho} \gamma^{n\frac{\rho_\gamma-\rho}{1+\rho}}
=e^{(1+\rho)\log\big(P^n_{\scalebox{1}{\tiny$\bm U$}}(\bm{u})^{\frac{1}{1+\rho_\gamma}}\big)}\gamma^{n\frac{\rho_\gamma-\rho}{1+\rho_\gamma}},\label{itw_kh3}
\end{align}
where we used $a^{(1+\rho)}=e^{(1+\rho)\log(a)}$.  Using \eqref{relation}  into \eqref{itw_kh1}, we may express $\gamma^n$ in terms of the $E_s(\cdot)$ function and its derivative $E_s^\prime(\cdot)$ as
\begin{align}
\gamma^n
=e^{E_s(\rho_\gamma,P^n_{\scalebox{.8}{\tiny$\bm U$}}(\bm{u}))-
(1+\rho_\gamma)E_s^\prime(\rho_\gamma,P^n_{\scalebox{.8}{\tiny$\bm U$}}(\bm{u}))},\label{itw_kh4}
\end{align}
Inserting the right hand side of \eqref{itw_kh4} into  \eqref{itw_kh3}, we obtain
\begin{align}
\min_{\lambda\geq 0}
\Bigg(\sum_{\bm u
}P^n_{\scalebox{1}{\tiny$\bm U$}}(\bm{u})^{\frac{1}{1+\rho}}
\biggl(\frac{\gamma^n}{P^n_{\bm U}(\bm u)}\biggr)^{\frac{(-1)^{i}\lambda}{1+\rho}}\Bigg)^{1+\rho}=
 e^{E_s(\rho_\gamma,P^n_{\scalebox{.8}{\tiny$\bm U$}}(\bm{u}))-
(\rho-\rho_\gamma)E_s^\prime(\rho_\gamma,P^n_{\scalebox{.8}{\tiny$\bm U$}}(\bm{u}))}.\label{itw_kh5}
\end{align}

Finally, combining \eqref{itw_kh1.5} and \eqref{itw_kh5} respectively for \eqref{negativelambda} and \eqref{positivelambda}, and using the definitions \eqref{Esi_itw_2018_1} and \eqref{Esi_itw_2018_2}, we conclude the proof.
\end{IEEEproof}

%

\begin{lemma}
\label{itw_lema_5} 
Let $E(\rho)$ be a function of $\rho$. The function $f_1(\gamma)=\max_{\rho\in[0,1]}E(\rho)-\Es[1]{}(\rho,P_{U},\gamma)$ is non-decreasing with respect to $\gamma$ and $f_2(\gamma)=\max_{\rho\in[0,1]}E(\rho)-\Es[2]{}(\rho,P_{U},\gamma)$ is non-increasing with respect to $\gamma$.
\end{lemma}
\begin{IEEEproof}
Let $\gamma,\gamma^\prime\in [0,1]$ where $\gamma\leq\gamma^\prime$, or equivalently $\frac{1}{1+\rho_\gamma}\leq\frac{1}{1+\rho_{\gamma^\prime}}$, where $\rho_\gamma$ is  defined in \eqref{11}.
Considering \eqref{Esi_itw_2018_1}  we conclude that  for all values of $\rho$ we have 
$\Es[1]{}(\rho,P_{U},\gamma)\geq \Es[1]{}(\rho,P_{U},\gamma^\prime)$. 
Thus, the maximum of $E(\rho)-\Es[1]{}(\rho,P_{U},\gamma)$ is not greater than the maximum of $E(\rho)- \Es[1]{}(\rho,P_{U},\gamma^\prime)$ meaning that $f_1(\gamma)\leq f_1(\gamma^\prime)$ or that $f_1(\gamma)$ is non-decreasing in $\gamma$.

Similarly, if $\gamma\leq\gamma^\prime$, by considering \eqref{Esi_itw_2018_2}  we conclude that  for all values of $\rho$ we have 
$\Es[2]{}(\rho,P_{U},\gamma)\leq \Es[2]{}(\rho,P_{U},\gamma^\prime)$. Using the same reasoning, we have $f_2(\gamma)\geq f_2(\gamma^\prime)$, or equivalently that $f_2(\gamma)$ is non-increasing in $\gamma$.
 
\end{IEEEproof}
\begin{lemma}
\label{itw_lema4}
Let $k_1(\gamma)$ and $k_2(\gamma)$ be respectively continuous non-decreasing and non-increasing functions with respect to $\gamma\in[0,1]$. The optimal $\gamma^\star$ maximizing $\min_{i=1,2} k_i(\gamma)$ satisfies the following equation
\begin{align}
k_1(\gamma^\star)=k_2(\gamma^\star).\label{itw_mi1}
\end{align}
When \eqref{itw_mi1} does not have any solution, we have $\gamma^\star=0$ if  $k_1(0)>k_2(0)$, and $\gamma^\star=1$ otherwise.
\end{lemma}
\begin{IEEEproof}
The relative behavior of a non-decreasing function with a non-increasing function can be categorized in three cases.
\begin{enumerate}
	\item We focus on the first case where $k_1(0)<k_2(0)$  and  $k_1(1)>k_2(1)$, 
i.e., there exists a $\gamma^\star$ such that $k_1(\gamma^\star)=k_2(\gamma^\star)$. In this case,  the function $\min_i k_i(\gamma)$  is non-decreasing from $[0,\gamma^\star)$, and non-increasing from $(\gamma^\star,1]$. Thus, the maximum over $\gamma$ of $\min_i k_i(\gamma)$ occurs at $\gamma=\gamma^\star$.

\item If $k_1(0)<k_2(0)$ and $k_1(1)<k_2(1)$, $k_1(\gamma)$ and $k_2(\gamma)$ do not cross in $\gamma\in[0,1]$. Hence,  we have $\min_i k_i(\gamma)=k_1(\gamma)$ and obviously since it is an non-decreasing function the maximum over $\gamma$ occurs at $\gamma=\gamma^\star=1$.

\item When  $k_1(0)\geq k_2(0)$, we have $\min_i k_i(\gamma)=k_2(\gamma)$ and hence $\gamma^\star=0$.

\end{enumerate}

\end{IEEEproof}

\begin{lemma}
\label{itw_lema_concavehull}
Let $L_{0s}(\rho)=L_0(\rho)-L_s(\rho)$ where $L_0(\rho)$ is a continuous function and $L_s(\rho)$ is a convex function of $\rho$. Then, 
\begin{align}
\bar L_{0s}(\rho)\leq\bar L_0(\rho)-L_s(\rho),\label{formula_itw_lema_concavehull}
\end{align}
where $\bar L_{0s}$ and $\bar L_{0}$  denote 
the concave hull of  $L_{0s}(\rho)$ and  $L_{0}(\rho)$, respectively.
\end{lemma}
\begin{IEEEproof}
From the definition of concave hull in \eqref{def_concave_huul_itw}, the left hand side of \eqref{formula_itw_lema_concavehull} is given by
\begin{align}
\bar L_{0s}(\rho)= \sup_{\substack{\rho_1,\rho_2,\theta\in[0,1] \,:
\\
\theta\rho_1+(1-\theta)\rho_2=\rho} }\Bigl\{
\theta L_{0s}(\rho_1)
+(1-\theta) L_{0s}(\rho_2)
\Bigr\}.
\label{itw_29_4_2018_10_09}
\end{align}
Using the definition of $L_{0s}(\rho)$, the right hand side of \eqref{itw_29_4_2018_10_09} is simplified as
\begin{align}
\theta L_{0s}(\rho_1)
+(1-\theta) L_{0s}(\rho_2)=
\theta L_{0}(\rho_1)
+(1-\theta) L_{0}(\rho_2)
-\theta L_{s}(\rho_1)
-(1-\theta) L_{s}(\rho_2).
\end{align}
Since $L_s(\rho)$ is a convex function of $\rho$, and so $\theta L_{s}(\rho_1)
+(1-\theta) L_{s}(\rho_2)\geq L_s(\theta \rho_1+(1-\theta)\rho_2) $, we further obtain that
\begin{align}
\theta L_{0s}(\rho_1)
+(1-\theta) L_{0s}(\rho_2)
\leq \theta L_{0}(\rho_1)
+(1-\theta) L_{0}(\rho_2)
-L_s(\rho),\label{itw_29_4_2018_11_32}
\end{align}
where we used that $\theta\rho_1+(1-\theta)\rho_2=\rho$. Taking supremum from both sides of  \eqref{itw_29_4_2018_11_32}, in view of \cite[Sec. 2.9]{goldberg}, we obtain that
\begin{align}
\sup_{\substack{\rho_1,\rho_2,\theta\in[0,1] \,:
\\
\theta\rho_1+(1-\theta)\rho_2=\rho} }\Bigl\{
\theta L_{0s}(\rho_1)
+(1-\theta) L_{0s}(\rho_2)
\Bigr\}\leq
\sup_{\substack{\rho_1,\rho_2,\theta\in[0,1] \,:
\\
\theta\rho_1+(1-\theta)\rho_2=\rho} }\Bigl\{
\theta L_{0}(\rho_1)
+(1-\theta) L_{0}(\rho_2)
\Bigr\}-L_s(\rho),\label{tamam_itw}
\end{align}
concluding the proof.
\end{IEEEproof}

\bibliographystyle{IEEEtran}	
\bibliography{IEEEabrv,gcitation}

\begin{thebibliography}{1}
\providecommand{\url}[1]{#1}
\csname url@samestyle\endcsname
\providecommand{\newblock}{\relax}
\providecommand{\bibinfo}[2]{#2}
\providecommand{\BIBentrySTDinterwordspacing}{\spaceskip=0pt\relax}
\providecommand{\BIBentryALTinterwordstretchfactor}{4}
\providecommand{\BIBentryALTinterwordspacing}{\spaceskip=\fontdimen2\font plus
\BIBentryALTinterwordstretchfactor\fontdimen3\font minus
  \fontdimen4\font\relax}
\providecommand{\BIBforeignlanguage}[2]{{%
\expandafter\ifx\csname l@#1\endcsname\relax
\typeout{** WARNING: IEEEtran.bst: No hyphenation pattern has been}%
\typeout{** loaded for the language `#1'. Using the pattern for}%
\typeout{** the default language instead.}%
\else
\language=\csname l@#1\endcsname
\fi
#2}}
\providecommand{\BIBdecl}{\relax}
\BIBdecl

\bibitem{gala}
R.~Gallager, \emph{Information Theory and Reliable Communication}.\hskip 1em
  plus 0.5em minus 0.4em\relax John Wiley \& Sons, 1968.

\bibitem{Cs2}
I.~Csisz{\'a}r, ``Joint source-channel error exponent,'' \emph{Probl. Control
  Inf. Theory}, vol.~9, no.~1, pp. 315--328, 1980.

\bibitem{zhong}
Y.~Zhong, F.~Alajaji, and L.~L. Campbell, ``On the joint source-channel coding
  error exponent for discrete memoryless systems,'' \emph{{IEEE} Trans. Inf.
  Theory}, vol.~52, no.~4, pp. 1450--1468, Apr. 2006.

\bibitem{jscc}
A.~Tauste~Campo, G.~Vazquez-Vilar, A.~Guill{\'e}n~i F{\`a}bregas, A.~Martinez,
  and T.~Koch, ``A derivation of the source-channel error exponent using
  nonidentical product distributions,'' \emph{{IEEE} Trans. Inf. Theory},
  vol.~60, no.~6, pp. 3209--3217, Jun. 2014.

\bibitem{farkas}
L.~Farkas and T.~Kói, ``Random access and source-channel coding error
  exponents for multiple access channels,'' \emph{{IEEE} Trans. Inf. Theory},
  vol.~61, no.~6, pp. 3029--3040, Jun. 2015.

\bibitem{isit2017}
A.~Rezazadeh, J.~Font-Segura, A.~Martinez, and A.~Guill{\'e}n~i F{\`a}bregas,
  ``An achievable error exponent for the multiple access channel with
  correlated sources,'' in \emph{IEEE Int. Symp. Inf. Theory (ISIT)}, 2017.

\bibitem{fans_minimax}
K.~Fan, ``Minimax theorems,'' \emph{Proc. of the Nat. Acad. Sci.}, vol.~39, pp.
  42--47, 1953.

\bibitem{polya}
H.~Polyanskiy, V.~Poor, and S.~Verd{\'u}, ``Channel coding rate in the finite
  blocklength regime,'' \emph{{IEEE} Trans. Inf. Theory}, vol.~56, no.~5, pp.
  2307--2359, May 2010.

\bibitem{goldberg}
R.~Goldberg, \emph{Methods of Real Analysis}.\hskip 1em plus 0.5em minus
  0.4em\relax Blaisdell Pub. Co., 1964.

\end{thebibliography}

\end{document}